\documentclass[12pt]{article}
\usepackage{amsmath,multirow}
\usepackage{amssymb}
\usepackage{graphicx}
\usepackage[comma]{natbib}
\usepackage[colorlinks,linkcolor=red,anchorcolor=blue,citecolor=blue]{hyperref}
\usepackage{float}
\usepackage{bm}
\usepackage{longtable}
\usepackage{multirow,booktabs}
\usepackage{lineno,hyperref}
\usepackage{rotating}
\usepackage{chngpage}
\usepackage{color, soul}
\usepackage{diagbox}
\usepackage{lscape}
\usepackage{enumerate}
\usepackage{dsfont}
\usepackage[justification=centering]{caption}
\newtheorem{theorem}{Theorem}

\newtheorem{remark}{Remark}

\newtheorem{definition}{Definition}
\newtheorem{lemma}{Lemma}
\newtheorem{assumption}{H.}
\makeatletter \def\@biblabel#1{#1.} \makeatother
\allowdisplaybreaks

\catcode`@=11 \@addtoreset{equation}{section} \catcode`@=12

\begin{document}
\title{Inference for multiple change-points in generalized integer-valued autoregressive model
\footnotetext{$^1$School of Mathematics and Statistics, Liaoning University, Shenyang, China\\
\indent~~$\ast$Corresponding author, E-mail: wangdehui@lnu.edu.cn\\
}
}
\author{Danshu Sheng$^1$, Dehui Wang$^{1*}$}
\date{}
\maketitle
\begin{center}
\begin{minipage}{14.5truecm}
{{\bf Abstract}~In this paper, we propose a computationally valid and theoretically justified methods, the likelihood ratio scan method (LRSM), for estimating multiple change-points in a piecewise stationary generalized conditional integer-valued autoregressive process.
LRSM with the usual window parameter $h$ is more satisfied to be used in long-time series with few and even change-points vs. LRSM with the multiple window parameter $h_{mix}$ performs well in short-time series with large and dense change-points.
The computational complexity of LRSM can be efficiently performed with order $O((\log n)^3 n)$.
Moreover, two bootstrap procedures, namely parametric and block bootstrap, are developed for constructing confidence intervals (CIs) for each of the change-points.
Simulation experiments and real data analysis show that
the LRSM and bootstrap procedures have excellent performance and are consistent with the theoretical analysis.}
\\
~\\
$\mathbf{Keywords:}$Piecewise stationary GCINAR process$\cdot$ Multiple change-points estimation$\cdot$  Likelihood ratio $\cdot$  Confidence interval $\cdot$ Count time series
\end{minipage}
\end{center}
\section{Introduction}
Modeling and analysis of non-stationary count time series have attracted a lot of attention over the past years.
Although complex non-stationary models have been developed in different fields, they are often difficult to explain.
The concept of piecewise stationary models has become a popular method by dividing nonstationary data into several stationary parts.
Among the different types of piecewise stationary models, the so-called multiple change-points (MCP) models have received special attention.
Studies of change-points models date back to \cite{Page1954,Page1955}. Since then, this topic has been of interest to statisticians and researchers in many other fields. 
So far, many excellent articles, such as \cite{Lee2003}, \cite{Davis2006}, \cite{Ngai2014}, \cite{Chen2021}, \cite{Aue2013}, \cite{Niu2016}, \cite{Casini2018}, \cite{Truong2020}, just to name a few, have studied and reviewed methodological issues related to estimation, detection and computation for continuous time series models involving structural changes.
In contrast, the research of count time series models involving change-points is still mainly focused on the change-points detection,
and there is little research on change-points estimation and computation.

One common and useful model to describe simple stationary count time series data is integer-valued autoregressive (INAR) time series model.
Since \cite{Al1987} proposed the INAR(1) model based on binomial thinning operator \citep{SV1979}, modeling INAR-type models based on different thinning operators have become a common approach and have been widely used in many fields like epidemiology, social sciences, economics, life sciences and others.
To make the integer-valued models more flexible for practical purposes, many scholars have extended the INAR (1) model by changing thinning operators, innovation.
For example, just to name a few, the INAR($p$) model based on generalized thinning operator proposed by \cite{Latour1998};
the mixture INAR(1) model based on the mixture of Pegram and thinning operators studied by \cite{Khoo2017};
the bounded binomial autoregressive (BAR) model proposed by \cite{McKenzie1985};
the nonlinear INAR(1) model, self-exciting threshold autoregressive process introduced by \cite{Monte2006}.
Actually, most of these INAR-type models can be written as the following $p$-order causal and stationary generalized conditional integer-valued autoregressive (GCINAR($p$)) process.
\begin{definition}\label{GINAR}
Considering a $\mathbb{N}_0$-valued ($\mathbb{N}_0=\mathbb{ N}\bigcup\{0\}$) GCINAR(p) process $\{X_{t}\}_{t\in \mathbb{Z}}$, where the conditional mean is defined by the following recursion
\vspace{-1mm}
\begin{align}\label{ginar}
{\rm E}(X_{t}|\mathcal{F}_{t-1})=\sum\limits_{k=1}^p\beta_{k} X_{t-k}+\beta_0.
\end{align}
$\mathcal{F}_{t}=\sigma(X_{s},s\leq t)$ is the $\sigma$-field generated by the whole information up to time $t$.
The parameter vector $\bm{\theta}=(\beta_0,...,\beta_p)^{\top}$ satisfied $\sum_{i=1}^p\beta_{i}<1$, $\beta_{i}\geq0$ for $i=1,...,p$ and $\beta_0>0$.
\end{definition}
\begin{remark}
Clearly, GCINAR can be called a conditional linear AR (CLAR, {\rm\cite{Grunwald2000}}) model in term of the form of the model.
However, the change-point of the continuous-type CLAR models has been studied by some scholars, see {\rm\cite{YauZhao2016,Ng2022}}.
Therefore, the focus of this paper is on the model defined on integer values and the conditional expectation is linear.
For example, the INAR-type model based on thinning operators, the INARCH models {\rm \citep{Wei2010}}, the BAR models {\rm\citep{McKenzie1985}} and among others.
\end{remark}
Then, a simple and useful idea to model piecewise stationary count time series is to construct the following GCINAR model with multiple change-points, that is, the so-called multiple change-points generalized conditional integer-valued autoregressive model (MCP-GCINAR).
\begin{definition}\label{Def1}
The MCP-GCINAR process with $m$ change-points $\{X_t\}_{t=1}^n$ is defined by the recursion:
\begin{align}\label{CGINAR}
{\rm E}(X_{t}|\mathcal{F}_{t-1})
=\begin{cases}
\beta_{1,1}X_{t-1,1}+...+\beta_{p_1,1}X_{t-p_1,1}+\beta_{0,1},~0<t\leq\tau_{1},\\
\vdots\\
\beta_{1,j} X_{t-\tau_{j-1}-1,j}+...+\beta_{p_j,j} X_{t-\tau_{j-1}-p_j,j}+\beta_{0,j},~\tau_{j-1}<t\leq\tau_{j},\\
\vdots\\
\beta_{1,m} X_{t-\tau_m-1,m}+...+\beta_{p_m,m} X_{t-\tau_m-p_m,m}+\beta_{0,m},\tau_{m}<t\leq n,\\
\end{cases}
\end{align}
where $\bm{\tau}=(\tau_1,\tau_2,...,\tau_{m})$ denotes the vector of unknown locations of change-points, $\tau_0=0$ and $\tau_{m+1}=n$.
Each change-point location $\tau_j$ is an integer between $1$ and $n-1$ inclusive, and the change-points are ordered such that $\tau_{j_1}< \tau_{j_2}$ if, and only if, $j_1<j_2$.
The time series vector $(X_1, X_2,..., X_n)$ can be written as \begin{align}\label{xn}
(X_{1,1},...,X_{n_1,1},...,X_{\tau_{j-1}+1,j},...,X_{\tau_{j}+n_j,j},...,X_{\tau_{m}+1,m+1},...,X_{\tau_{m}+n_{m+1},m+1}),
\end{align}
where $n_j=\tau_j-\tau_{j-1}$ for $j = 1,...,m+1$ and $n = n_1 + n_2 +...+ n_{m+1}$.
In particular, the $j$th piece $\{X_{t,j}\}_{t=1}^{n_j}$ of the series is modeled as a stationary GCINAR process (\ref{ginar}) with order $p_j$,
\begin{align*}
{\rm E}(X_{t,j}|\mathcal{F}_{t-1})=\beta_{1,j} X_{t-1,j}+...+\beta_{p_j,j} X_{t-p_j,j}+\beta_{0,j},
\end{align*}
$\bm{\theta_j}=(\beta_{0,j},\beta_{1,j},...,\beta_{p_j,j})^{\top}$ is the parameter vector corresponding to the $j$th segment GCINAR($p_j$) process, which is assumed to be an interior point of the compact space $\bm{\Theta}_j(p_j)=[\delta,1/\delta]\times[0,1-\delta]^{p_j}\cap\mathbb{M}_j$, where $\mathbb{M}_j=\{0\leq\sum_{k=1}^{p_j}\beta_{k,j}\leq1-\delta<1\}$, $\delta\in(0,1)$ is a constant.
\end{definition}

So far, most research has focused on the change-points detection of the INAR-type models.
For example, among others,
\cite{Kang2009}, \cite{YuandKim2020} and \cite{Lee2022} considered the problem of testing for a parameter change in different types of INAR(1) models by taking advantage of the cumulative sum (CUSUM) test.
\cite{Chattopadhyay2021} considered the problem of change-point analysis for the INAR(1) model with time-varying covariates.
\cite{Yu2022} applied the empirical likelihood ratio (ELR) test to uncover a structural change in INAR processes.
\cite{Wei2007,Wei2009a,Wei2009b,WeiTes2009} studied the Shewhart, combined jumps, exponentially weighted moving average and cumulative sum charts for controlling the Poisson counts process.

However, few research has studied the estimation of change-points, especially optimization.
\cite{DiopKengne2021b} derive a data-driven procedure based on the slope heuristic to calibrate the penalty term of the contrast to achieve general integer-valued time series change-points estimation, and optimized by dynamic programming (DP) algorithm;
\cite{Sheng2023} studied the change-points analysis of the MCP-GCINAR model based on minimum description length (MDL) principle, and optimized by genetic algorithm (GA).
Although these two algorithms have been suggested for implementing the inference of change-points in the MCP-GCINAR model, optimization can be computationally very expensive because the number of possible change-points combinations grows exponentially as the sample size grows.
Specifically, the GA involves various tuning parameters, and the DP algorithm exhibits a computational order $O(n^2)$.
In fact, if the computational complexity of optimal methods is too great for the
application at hand, we can resort to an ``approximate methods".
That is, we can roughly judge a potential change-points set which is far smaller than the sample size by some methods,
and then the optimization problem based on information criteria is realized by selecting the best subset of this potential change-points set.
Of course, it is necessary to ensure that the number of elements in the potential change-points is higher than the true number of change-points,
and there are some subsets in a neighborhood of all true change-points.
This ``approximate methods" approach is also often considered in some continuous-type models, such as
\cite{YauZhao2016}, \cite{Ng2022}, \cite{Ngai2014}, \cite{Safikhani2022}.
In addition, it should be noted that the generalized likelihood ratio scan method (GLRSM) proposed by \cite{Ng2022} requires the likelihood function to be a measurable and continuous function with respect to $\{X_t\}$.
And the proof for the GLRSM relies on the application of the continuous mapping theorem.
Consequently, it can not be simply applied to count time series models.

Inspired by the above review/discussion, we propose two computationally valid and theoretically justified methods for off-line change-points inference in the MCP-GCINAR processes.
The main contributions in this paper are as follows:
\begin{itemize}
 \item We propose a computationally efficient and theoretically sound procedure LRSM for the inference of the MCP-GCINAR model in the case of long-time series
with few and even change-points.
Performing LRSM involves three steps:
first, a likelihood ratio scan statistic is used to obtain a potential change-points set;
second, a model selection procedure based on the MDL principle is employed to give a set of consistent change-points estimates;
and finally, an exhaustive method is used to determine the final change-points in an extended local window and its convergence and asymptotic distribution are given.
Also, we demonstrated that the computational order of the LRSM can be as low as $O((\log n)^3 n)$.
After simulation results and real data analysis, LRSM performs well in samples of long-time series with few and even change-points.
  \item We construct the CIs based on two bootstrap procedures, parametric and block bootstrap, for each estimated change-point.
  In addition, the validity of the two bootstrap procedures are discussed.
\end{itemize}
Although these methods are similar to the problem of change-points estimation in continuous-type models,
there are many differences  in the derivation of asymptotic properties, model assumptions, and algorithm implementation.
For example, the techniques and assumptions used to develop the asymptotic properties in LRSM,
the techniques used to develop the validity of the two bootstrap procedures.

The rest contents of this article are organized as follows.
In Section 2, we state the details of the three-step LRSM and construct the confidence intervals by two bootstrap procedures.
Section 3 discusses the tuning parameters, computational complexity and some implementation issues.
Extensive simulation studies and real data applications are given in Sections 4 and 5 respectively.
The article ends with a conclusion section.
All proofs, some implementation algorithms and additional simulation results are given in the Appendix.

\section{MCP-GCINAR model change-points inference based on the three-step LRSM}\label{Set2}
In this section, we state the three-step LRSM to implement the inference of the MCP-GCINAR model.
Following Section 2.1 in \cite{Sheng2023}, considering the complexity of proof and computation, Poisson quasi-maximum likelihood (PQML) estimation as the cost function for the MDL criterion is reasonable and convenient.
Thus, we first review the Poisson quasi-maximum log-likelihood for the GCINAR process and provide some notations and assumptions used in the LRSM.

The conditional Poisson quasi-maximum log-likelihood for the data set is obtained by
\begin{align}\label{PQMLeq}
L_{n}(k,\bm{\theta},p)&=\sum\limits_{t=k+1}^{n+k}\left[X_{t}\log \xi_{t}(\bm{\theta},p|{\bm X_{t-1}})-\xi_{t}(\bm{\theta},p|{\bm X_{t-1}})\right]\\
&=\sum\limits_{t=k+1}^{n+k}\ell_t(\bm{\theta},p),\nonumber
\end{align}
where $\{X_{t}\}_{t=k+1-p}^{n+k}$ is generated from the GCINAR process (\ref{ginar}), $k\in \mathds{Z}$ and $k\geq p$,
$\xi_{t}(\bm{\theta},p|{\bm X}_{t-1})$ $=\sum\limits_{i=1}^{p}\beta_{i} X_{t-i}+\beta_{0}\triangleq{\bm X}^{\top}_{t-1}\bm{\theta}$ with ${\bm X}_{t-1}=(1, X_{t-1},X_{t-2},...,X_{t-p})^{\top}$ and $\bm{\theta}=(\beta_{0},\beta_{1},...,\beta_{p})^{\top}$.
Then the PQML estimate of $\bm{\theta}$ is defined by
\begin{align*}
{\bm{\hat\theta}}=\arg\max\limits_{\bm{\theta}\in{\bm\Theta}(p)} L_{n}(k,\bm{\theta},p).
\end{align*}
Assuming that $0<\sum\limits_{i=1}^{p}\beta_{i}<1$ and ${\rm E}(X_{t})<\infty$, then ${\bm{\hat\theta}}$ satisfy the following asymptotic normality
\begin{align*}
\sqrt{n}(\bm{\hat{\theta}}-\bm{\theta}^0)\xrightarrow{d} N(0,\bm{\Sigma})~~~\text{as~~~$n\rightarrow \infty$},
\end{align*}
where $\bm{\theta}^0$ denote the true parameter value of $\bm{\theta}$. The asymptotic variance matrix of the PQML estimate can be consistently estimated by $\hat{\bm{\Sigma}}_n=\hat{\bm{J}}_n^{-1}(k,\hat{\bm{\theta}})\hat{\bm{I}}_n(k,\hat{\bm{\theta}})\hat{\bm{J}}_n^{-1}(k,\hat{\bm{\theta}})$ with
\begin{align}
\hat{\bm{J}}_n(k,\hat{\bm{\theta}})&=\frac{1}{n}\sum\limits_{t=k+1}^{n+k}\left.\frac{1}{\xi_{t}(\bm{\theta},p|{\bm X_{t-1}})}\frac{\partial\xi_{t}(\bm{\theta},p|{\bm X_{t-1}})}{\partial \bm{\theta}}\frac{\partial\xi_{t}(\bm{\theta},p|{\bm X_{t-1}})}{\partial \bm{\theta}^{\mathrm{T}}}\right|_{\bm{\theta}=\hat{\bm{\theta}}},\label{SE1}\\
\hat{\bm{I}}_n(k,\hat{\bm{\theta}})&=\frac{1}{n}\sum\limits_{t=k+1}^{n+k}\left.(\frac{X_{t}}{\xi_{t}(\bm{\theta},p|{\bm X_{t-1}})}-1)^2\frac{\partial\xi_{t}(\bm{\theta},p|{\bm X_{t-1}})}{\partial \bm{\theta}}\frac{\partial\xi_{t}(\bm{\theta},p|{\bm X_{t-1}})}{\partial \bm{\theta}^{\mathrm{T}}}\right|_{\bm{\theta}=\hat{\bm{\theta}}}.\label{SE2}
\end{align}
~\\
$\emph{\textbf{Notations:}}$
\begin{itemize}
 \item Denote this whole class of the MCP-GCINAR model by $\mathcal{M}$ and any model from this class by $\mathcal{F} \in\mathcal{ M}$.
 \item Upper bounds for the GCINAR orders $p$ are represented by $p_{\max}$. Setting $\bm{p}=(p_1,...,p_{m+1})$ belongs to the parameter domain $\mathcal{P}=(0,p_{\max}]^{m+1}\bigcap \mathds{Z}^{m+1}$, ${\bm\theta}(m)=(\bm{\theta_1},...,\bm{\theta_{m+1}})$ belongs to the parameter domain $\bm{\Theta}=\prod\limits_{j=1}^{m+1}\bm{\Theta}_j(p_j)$.
 \item Denote the true number of change-points by $m_0$, the set of true change-points by $\mathcal{J}_0=\{\tau_1^0,\tau_2^0,...,\tau_{m_0}^0\}$, the set of true order by ${\bm{p}}_0=\{p_1^0,...,p_{m+1}^0\}$.
 \item Let $|\mathcal{J}|$ be the cardinality of the set $\mathcal{J}$.
\end{itemize}
$\emph{\textbf{Assumptions:}}$
\begin{assumption}\label{H1}
For each segment, assume that $\bm{\theta}_j$ is an interior point of the compact space $\bm{\Theta}_j(p_j)$ and satisfies $\sum_{i=1}^{p_j}\beta_{i,j}<1$, $\beta_i\geq0$ for $i=1,...,p_j$.
\end{assumption}
\begin{assumption}\label{H2}
Assume that there exists a $\epsilon_{\theta}>0$ such that $\min_{1\leq j\leq m_0}||\bm{\theta}_{j+1}- \bm{\theta}_{j}||>\epsilon_{\theta}$.
\end{assumption}
\begin{assumption}\label{H3}
Assume that there exists a $\epsilon_{\tau}>0$ such that $\min_{0\leq j\leq m_0}|\tau_{j+1}-\tau_{j}|>n\epsilon_{\tau}$.
\end{assumption}
\begin{assumption}\label{H7}
For each segment, assume that there exist a constant $\epsilon_K>0$ such that
${\rm E}\big[e^{|l_{t}(\bm{\theta}_j)-{\rm E}[l_{t}(\bm{\theta}_j)]|}\big]<\epsilon_K$ for all $\bm{\theta}_j\in\bm{\Theta}_j(p_j)$.
\end{assumption}
\begin{assumption}\label{H4}
For each segment, assume that ${\rm E}|X_{t}|^{4+\epsilon_X}<\infty$ for some $\epsilon_X>0$.\\
\end{assumption}
\vspace{-5mm}

Under the assumption H.\ref{H1}, the $j$th segment process $\{X_{t,j}\}$ is ergodic and has a strictly stationary solution, additionally, $\{X_{t,j}\}$ is strong mixing with geometric rate.
Assumption H.\ref{H2} imposes restrictions on the parametric differences between each segment, which is essential for the existence of change-points.
Assumption H.\ref{H3} imposes restrictions on the distances between change-points. To accurately estimate the specified GCINAR parameter values,  the segments must have a sufficient number of observations. If not, the estimation is over-determined and the likelihood has an infinite value.
This assumption is common in likelihood-based model selection, such as \cite{Davis2006}, \cite{DavisYau2013}, \cite{YauZhao2016}.
Assumption H.\ref{H4} is proposed to guarantee the consistency of the change-points estimate in Theorem \ref{LRSM2_theory}.
Assumption H.\ref{H7} provides the condition for the establishment of the large deviation conclusion in Theorem \ref{LRSM1_theory}.
It's worth mentioning that, if the arbitrary moments of subsegment model are bounded, assumption H.\ref{H4} and H.\ref{H7} are unnecessary.
For example, if the MCP-GCINAR model represent the BAR model \citep{McKenzie1985} with multiple change-points,
it is no longer necessary to assume H.\ref{H4} and H.\ref{H7}.
\subsection{First Step: obtain potential change-points based on LR scan statistics}\label{section2.1}

For $t=h,..., n-h,$ define the scanning window at $t$ and the corresponding observations as
\begin{align*}
W_t(h)=\{t-h+1,...,t+h\}~~~\text{and}~~~X_{W_t(h)}=(X_{t-h+1},...,X_{t+h}),
\end{align*}
respectively, where $h$ is called the window radius. To establish asymptotic theory, we assume that $h = h(n)$ depends on the sample size $n$. Then the likelihood ratio scan statistic for the scanning window $W_t(h)$ by
\begin{align*}
S_h(t)=\frac{1}{h}L_{h}(t-h,\hat{\bm{\theta}}_1,p_1)+\frac{1}{h}L_{h}(t,\hat{\bm{\theta}}_2,p_2)-\frac{1}{h}L_{2h}(t-h,\hat{\bm{\theta}},p)
\end{align*}

where $L_{h}(t-h,\hat{\bm{\theta}}_1,p_1), L_{h}(t,\hat{\bm{\theta}}_2,p_2), L_{2h}(t-h,\hat{\bm{\theta}},p)$, defined by (\ref{PQMLeq}), are the Poisson quasi-likelihoods formed by the observations $\{X_s\}_{s=t-h+1}^t$, $\{X_s\}_{s=t+1}^{t+h}$, $\{X_s\}_{s=t-h+1}^{t+h}$, evaluated at the PQML estimates $\hat{\bm{\theta}}_1, \hat{\bm{\theta}}_2, \hat{\bm{\theta}}$, respectively.
Similar to AR model, each GCINAR($p_j$) model can be regarded as an GCINAR($p_{\max}$) model (with the last few coefficients equal to $0$), we can regard each GCINAR model as $p_{\max}$ order when deducing the theoretical results in this step.

Next, $S_h(t)$ scans the observed time series to obtain a sequence of likelihood ratio scan statistics $(S_h(h), S_h(h+1),..., S_h(n-h))$. As discussed by \cite{YauZhao2016}, after this construction, $S_h(t)$ tends to be larger if $t$ is change-point. In particular, if $h$ is selected as $2h<n\epsilon_{\tau}$ and $h>p_{\max}$, there is at most one change-point in each scanning window. Thus, we can derive a set of potential change points from local change point estimates, given by
\begin{align*}
\hat{\mathcal{J}}^{(1)}(h)=\Big\{\tau\in\{h,h+1,...,n-h\}: S_h(\tau)=\max\limits_{t\in(\tau-h,\tau+h]}S_h(t)\Big\},
\end{align*}
where $S_h(t)\triangleq0$, for $t<h$ and $t>n-h$. Clearly, if $S_h(\tau)$ is the maximum on the window $[\tau-h+1, \tau+h]$ centered on $\tau$, then $\tau$ is a local point of change estimate.

Denote the number of elements in $\hat{\mathcal{J}}^{(1)}(h)$ by $\hat{m}^{(1)}$, the local change-points estimates set by $\hat{\mathcal{J}}^{(1)}(h)=\{\tau_1^{(1)},\tau_2^{(1)},...,\tau_{\hat{m}^{(1)}}^{(1)}\}$. Theorem \ref{LRSM1_theory} states that all change-points can be identified in an $h$-neighborhood of $\hat{\mathcal{J}}^{(1)}(h)$ obtained in this step.
\begin{theorem}\label{LRSM1_theory}
Suppose Assumptions H.\ref{H1}-H.\ref{H7} hold, $2h<n\epsilon_{\tau}$ and $\epsilon_{\tau}>c$ for some $c>0$, then there exists some $d>0$ such that, for $h\geq d(log n)^3$,
\begin{align*}
{\rm P}\Big(\max\limits_{\tau\in\mathcal{J}_0}\min\limits_{s=1,...,\hat{m}^{(1)}}|\tau-\hat{\tau}_s^{(1)}|<h\Big)\rightarrow1.
\end{align*}
\end{theorem}

For the first step, make the following supplementaries through Remark \ref{LRSM1_remark}.
\begin{remark}\label{LRSM1_remark}
~\\
\vspace{-7mm}
\begin{itemize}
 \item To enhance practical performance, when we evaluate $S_h(t)$, the order $p$ for the three estimates $\hat{\bm{\theta}}_1$, $\hat{\bm{\theta}}_2$, $\hat{\bm{\theta}}$ can be individually selected by using the information criteria: Akaike information criterion (AIC) or Bayesian information criterion (BIC).
 \item Usually, $\hat{m}^{(1)}$ is much larger than the true number of change-points, especially if $h$ is small. In order to enhance computational efficiency, we can select the first $m_{\max}$ elements in $\hat{\mathcal{J}}^{(1)}$ that have the largest $S_h(t)$s as the final $\hat{\mathcal{J}}^{(1)}$, where $m_{\max}$ is a reasonably large number upper bound of the number of change points, such as $20$ or $50$.
\end{itemize}
Note that these two operations are only intended to enhance practical performance and computational efficiency, and are not necessary, especially the value of $m_{\max}$ has little effect on the estimation results.
\end{remark}

\subsection{Second Step: consistency estimation based on MDL principle}
The set of potential change-points $\hat{\mathcal{J}}^{(1)}$ can be viewed as a rough estimate of the change-points, which usually overestimates the true set of change-points. To detect the true change-points, the best subset of $\hat{\mathcal{J}}^{(1)}$ can be selected according to some prescribed information criterion (IC), such as BIC \citep{Yao1988}, MDL \citep{Davis2006}.
Here, MDL is selected as IC to select the most concise model as the optimal model.

Given a set of change-points $\mathcal{J}=\{\tau_1,...,\tau_m\}$, the MDL criterion is defined as
\begin{align}\label{MDL_eq}
\textbf{MDL}(m,\mathcal{J},\bm{p})=&\log(m)+(m+1)\log n+\sum\limits_{j=1}^{m+1}\log(p_j)+\sum\limits_{j=1}^{m+1}\frac{p_j+1}{2}\log(n_j)\nonumber\\
&-\sum\limits_{j=1}^{m+1}L_{n_j}(\tau_{j-1},\bm{\theta_j},p_j;\bm{X_j}).
\end{align}
where $\bm{X_j}=\{X_{\tau_{j-1}+1},...,X_{\tau_j}\}$.  
Given the potential change-points estimates $\hat{\mathcal{J}}^{(1)}$, the change-points and corresponding GCINAR orders can be estimated by
\begin{align}\label{MDL}
(\hat{m}^{(2)},\hat{\mathcal{J}}^{(2)},\bm{\hat{p}}^{(2)})=\arg\min\limits_{
m=|\mathcal{J}|,|\mathcal{J}|\subseteq \hat{\mathcal{J}}^{(1)},{\bm{p}\in\mathcal{P}}}
\textbf{MDL}(m,\mathcal{J},\bm{p}),
\end{align}
where $\hat{m}^{(2)}=|\hat{\mathcal{J}}^{(2)}|$, $\hat{\mathcal{J}}^{(2)}=\{\tau_1^{(2)},\tau_2^{(2)},...,\tau_{\hat{m}^{(2)}}^{(2)}\}$, $\bm{\hat{p}}^{(2)}=(\hat{p}_1^{(2)},...,\hat{p}_{\hat{m}^{(2)}}^{(2)})$.
Minimization MDL equation (\ref{MDL}) can be achieved by optimal partitioning (OP) algorithm \citep{Jackson2005}.
The implementation details of OP algorithm are given in Appendix.
Following \cite{Sheng2023}, which proved the consistency of estimates based on the MDL principle, we have the following Theorem \ref{LRSM2_theory}.
\begin{theorem}\label{LRSM2_theory}
Under the setting in Theorem \ref{LRSM1_theory} and suppose Assumption \ref{H4} hold, there is $\hat{m}^{(2)}\xrightarrow{p} m_0$. In addition, given that $\hat{m}^{(2)}= m_0$, we have
\begin{align*}
&{\rm P}\Big(\max\limits_{j=1,...,m_0}|\hat{\tau}_j^{(2)}-\tau_j^{0}|<h\Big)\rightarrow1,~~~\text{and}~~~\max\limits_{j=1,...,m_0}|\hat{p}_j^{(2)}-p_j^{0}|\xrightarrow{p}0.
\end{align*}
\end{theorem}
\subsection{Third Step: final change-points estimates}
Although the consistent estimate of $m$ can be achieved based on the MDL criterion, it should be noted that $\hat{\mathcal{J}}^{(2)}$ is a subset of $\hat{\mathcal{J}}^{(1)}$, Theorem \ref{LRSM2_theory} only implies that $\max_{j=1,...,m_0}|\hat{\tau}_j^{(2)}-\tau_j^{0}|=O_p(h)$, which is not optimal compared with the typical rate of $O_p(1)$.
Nevertheless, the Theorem \ref{LRSM2_theory} also guarantees that the true change-point $\tau_j^0$ is within $(\hat{\tau}_j^{(2)}-h,\hat{\tau}_j^{(2)}+h]$ with probability approaching 1 for all $j=1,...,\hat{m}^{(2)}$.
A simple idea is to use exhaustive methods to determine the final change-points in an extended local window $(\hat{\tau}_j^{(2)}-2h,\hat{\tau}_j^{(2)}+2h]$ and corresponding observations.
\cite{YauZhao2016} discusses in detail that this is possible, and if $3h<n\epsilon_{\tau}$, there is only one change point inside the extended local window.

Define the extended local window and corresponding observations for the $j$th estimated change-point $\hat{\tau}_j^{(2)}$ by
\begin{align*}
EW_j(h)=\{\hat{\tau}_j^{(2)}-2h+1,...,\hat{\tau}_j^{(2)}+2h\} ~~~\text{and}~~~X_{EW_j(h)}=\{X_{\hat{\tau}_j^{(2)}-2h+1},...,X_{\hat{\tau}_j^{(2)}+2h}\}.
\end{align*}
For $j=1,...,\hat{m}^{(2)}$, Let
\begin{align*}
L_j(\tau,\bm{\theta}_j,\bm{\theta}_{j+1})=L_{\tau-\hat{\tau}_j^{(2)}+2h}(\hat{\tau}_j^{(2)}-2h,\bm{\theta}_{j},\hat{p}_j^{(2)})+L_{\hat{\tau}_j^{(2)}+2h-\tau}(\tau,\bm{\theta}_{j+1},\hat{p}_{j+1}^{(2)}),
\end{align*}
where $L_{\tau-\hat{\tau}_j^{(2)}+2h}(\hat{\tau}_j^{(2)}-2h,\bm{\theta}_j,\hat{p}_j^{(2)})$ and $L_{\hat{\tau}_j^{(2)}+2h-\tau}(\tau,\bm{\theta}_{j+1},\hat{p}_{j+1}^{(2)})$ are defined in (\ref{PQMLeq}), i.e.,
\begin{align*}
&L_{\tau-\hat{\tau}_j^{(2)}+2h}(\hat{\tau}_j^{(2)}-2h,\bm{\theta}_j,\hat{p}_j^{(2)})=\sum\limits_{t=\hat{\tau}_j^{(2)}-2h+1}^{\tau}\ell_t(\bm{\theta}_j,\hat{p}_j^{(2)}),\\
&L_{\hat{\tau}_j^{(2)}+2h-\tau}(\tau,\bm{\theta}_{j+1},\hat{p}_{j+1}^{(2)})=\sum\limits_{t=\tau+1}^{\hat{\tau}_j^{(2)}+2h}\ell_t(\bm{\theta}_{j+1},\hat{p}_{j+1}^{(2)}).
\end{align*}
Define the final estimate as
\begin{align*}
\hat{\tau}_j^{(3)}=\arg\max\limits_{\tau\in(\hat{\tau}_j^{(2)}-h,\hat{\tau}_j^{(2)}+h]}L_j(\tau,\hat{\bm{\theta}}_j,\hat{\bm{\theta}}_{j+1}),
\end{align*}
where $\hat{\bm{\theta}}_j$ and $\hat{\bm{\theta}}_{j+1}$ are the maximizers of $L_{\tau-\hat{\tau}_j^{(2)}+2h}(\hat{\tau}_j^{(2)}-2h,\bm{\theta}_j,\hat{p}_j^{(2)})$ and $L_{\hat{\tau}_j^{(2)}+2h-\tau}(\tau,\bm{\theta}_{j+1},\hat{p}_{j+1}^{(2)})$ on $\bm{\Theta}_j(\hat{p}_j)$ and $\bm{\Theta}_j(\hat{p}_{j+1})$, respectively.
Following Theorem 1 and 2 in \cite{Cui2021}, we have the convergence and asymptotic distribution of the final estimates.
\begin{theorem}\label{LRSM3_theory}
Under the setting in Theorem \ref{LRSM2_theory} and $3h<n\epsilon_{\tau}$,
Then, we have
\begin{align*}
\hat{\tau}_j^{(3)}-\tau_j^0\xrightarrow{d}\arg\max\limits_{\tau}W_{j,\tau},
\end{align*}
where
\begin{align}\label{Wtau}
W_{j,\tau}=\left\{ \begin{array}{lr}
\sum\limits_{t=\tau_j^{0}+1}^{\tau_j^{0}+\tau}\left[\ell_t(\bm{\theta}_j^0,p_j^{0})-\ell_t(\bm{\theta}_{j+1}^0,p_{j+1}^{0})\right]&\tau>0,\\
0&\tau=0,\\
\sum\limits_{t=\tau_j^{0}+\tau}^{\tau_j^{0}-1}\left[\ell_t(\bm{\theta}_{j+1}^0,p_{j+1}^{0})-\ell_t(\bm{\theta}_j^0,p_j^{0})\right]&\tau<0
\end{array}
\right.
\end{align}
is a double-sided random walk. In particular, $\hat{\tau}_j^{(3)}=\tau_j^{0}+O_p(1).$
\end{theorem}

\subsection{CIs' approximation for the final change-points estimates.}
Although Theorem \ref{LRSM3_theory} deduces the asymptotic distribution $\arg\max\limits_{\tau}W_{j,\tau}$ of $(\hat{\tau}_j^{(3)}-\tau_j^0)$, it is difficult to use in practice.
According to the arguments of \cite{Cui2021} about the approximation distribution of $\arg\max\limits_{\tau}W_{j,\tau}$, we have the following theorem when change is small.
\begin{theorem}\label{asym}
Let $\bm{d}_j=\bm{\theta}_j^0-\bm{\theta}_{j+1}^0$, under the setting of Theorem \ref{LRSM3_theory}, if $||\bm{d}_j||\rightarrow0$ as $n\rightarrow0$, then
\begin{align}\label{CI_appromix}
\hat{\Delta}_j^{-1}(\hat{\tau}_j^{(3)}-\tau_j^0)\xrightarrow{d}\arg\max\limits_{r\in \mathbb{R}} \mathcal{B}(r)-\frac{1}{2}|r|,
\end{align}
where
$$\hat{\Delta}_j=(\hat{\bm{d}}_j^{\top}\hat{\bm{J}}_{j}\hat{\bm{d}}_j)^{-2}(\hat{\bm{d}}_j^{\top} \hat{\bm{I}}_{j}\hat{\bm{d}}_j) ~~~\text{and}~~~\hat{\bm{d}_j}=\hat{\bm{\theta}}_j-\hat{\bm{\theta}}_{j+1}.$$
With $\hat{\bm{J}}_{j}=\hat{\bm{J}}_{4h}(\hat{\tau}_j^{(3)}-2h,\hat{\bm{\theta}}_{j+1}^{(3)})$ and $\hat{\bm{I}}_{j}=\hat{\bm{I}}_{4h}(\hat{\tau}_j^{(3)}-2h,\hat{\bm{\theta}}_{j+1}^{(3)})$ are defined in (\ref{SE1}) and (\ref{SE2}).
$\mathcal{B}(r)$ is the two-sided standard Brownian motion in $\mathbb{R}$.
\end{theorem}
\begin{remark}
If the orders of GCINAR in two consecutive segments, denoted by $\hat{p}_j$ and $\hat{p}_{j+1}$, are not equal, then the GCINAR
parameter vectors $\hat{\bm{\theta}}_j$ and $\hat{\bm{\theta}}_{j+1}$ will be interpreted as being in the order of $\max(\hat{p}_j, \hat{p}_{j+1})$.
\end{remark}
To simplify notation, denote $V=\arg\max\limits_{r\in \mathbb{R}} \mathcal{B}(r)-\dfrac{1}{2}|r|$. The distribution of $V$ has been studied by \cite{Yao1987}, who demonstrates that $V$ has a symmetric distribution, and for $a > 0$, the distribution function is given by
\begin{align*}
P(V\leq a)=1+\sqrt{\frac{a}{2\pi}}\exp(-\frac{a}{8})+\frac{3}{2}\exp(a)\Phi(-\frac{3}{2}\sqrt{a})-\frac{1}{2}(a+5)\Phi(-\frac{1}{2}\sqrt{a}),
\end{align*}
where $\Phi(\cdot)$ represents the standard normal distribution function. Let $F_{\alpha/2}$ be the $(1-\alpha/2)$th quantile of $V$, that is $P(V\leq F_{\alpha/2})=1-\alpha/2$, and $F_{0.05}=7.6873$. 
Then an approximate $100(1-\alpha)\%$ confidence interval (CI) for $\tau_j^0$ can be constructed by
\begin{align}\label{CI}
CI={\Big[}\hat{\tau}_j^{(3)}-\lfloor \hat{\Delta}_jF_{\alpha/2}\rfloor-1,\hat{\tau}_j^{(3)}+\lfloor \hat{\Delta}_jF_{\alpha/2}\rfloor+1{\Big]},
\end{align}
where $\lfloor a\rfloor$ is the largest integer not greater than $a$.
Since the minimum distance between  change-points is much larger than the maximum
window radius $h$, i.e. $n\epsilon_{\tau}/h\rightarrow\infty$, the distance between the extended local windows $EW_j(h)$s diverge to $\infty$.
Under H.\ref{H1}, the GCINAR($p$) process is strongly mixing, the CIs constructed are asymptotically independent.
According to a Bonferroni-type argument, one can construct an asymptotically correct $1-\alpha$ simultaneous CI covering all $\hat{m}^{(2)}$ change-points by using a collection of $(1-\alpha)^{1/\hat{m}^{(2)}}$ CIs for each of a set of $\hat{m}^{(2)}$ change-points.
\subsection{Bootstrap approximation}\label{sect2.5}
Due to the restriction of condition $||\bm{d}_j||\rightarrow0$ in Theorem \ref{asym},
some scholars have confirmed the fact that the asymptotic theory of change-points estimates given by Theorem 3 in \cite{Cui2021} provides a poor approximation to the actual multimodal finite sample distribution under small parameter changes,
thus the asymptotic theory of change-points estimates given by Theorem \ref{asym} also has this problem.
Furthermore, the pivotal approximations in CI (\ref{CI}) work unsatisfactorily under medium and large parameter changes.
Considering these issues, inspired by \cite{Ng2022}, we propose the following two bootstrap procedures, parametric bootstrap and block bootstrap, to construct the CI for the change-point $\tau_j^0$.

{\bf\emph{Parametric Bootstrap Algorithm (PBA)}}. The basic idea of parametric bootstrap is to first use the model based on the PQML estimate to simulate replicated samples of the series before and after the change-point.
Then the bootstrap samples are used to approximate the asymptotic distribution of $\arg\max\limits_{\tau}W_{j,\tau}$ in Theorem \ref{LRSM3_theory}.
The detail of PBA for $\tau_j^{0}$ is given introduction in the following.
\begin{table}[H]
\tabcolsep0.02in
		\begin{tabular}{ll}
			\toprule
\multicolumn{2}{l}{\bf{Parametric Bootstrap Algorithm}:}\\
\midrule
{Input}:& The significance level $\alpha$, resample size $n_p$ and the sampling number $B$.\\
&The order estimates $\hat{p}_j^{(2)}$ and $\hat{p}_{j+1}^{(2)}$ from ``Second step".\\
&The parameter estimates $\hat{\bm{\theta}}_j$ and $\hat{\bm{\theta}}_{j+1}$, change-point estimate $\hat{\tau}_j^{(3)}$ from ``Third step".\\

\midrule
\multicolumn{2}{l}{\emph{Simulate replicated samples:}}\\
\cmidrule(lr){1-2}
\multicolumn{2}{l}{for $s=1:B$}\\
\multicolumn{2}{l}{~~~ simulate time series sample $\{\widetilde{X}_t^{(s)}\}_{t=1}^{n_p+1}$ follow the special GCINAR($\hat{\bm{\theta}}_j$,$\hat{p}_j^{(2)}$) model,}\\
\multicolumn{2}{l}{~~~ simulate time series sample $\{\widetilde{X}_t^{(s)}\}_{t=n_p+2}^{2n_p+1}$ follow the special GCINAR($\hat{\bm{\theta}}_{j+1}$,$\hat{p}_{j+1}^{(2)}$) model,}\\
\multicolumn{2}{l}{~~~ join them to form the resampled process $\{\widetilde{X}_t^{(s)}\}_{t=1}^{2n_p+1}$.}\\
\multicolumn{2}{l}{end}\\

\midrule
\multicolumn{2}{l}{\emph{Approximate the asymptotic distribution of} $\arg\max\limits_{\tau}W_{j,\tau}$:}\\
\cmidrule(lr){1-2}
\multicolumn{2}{l}{for $s=1:B$}\\
\multicolumn{2}{l}{~~~ obtain $\{\widetilde{\ell}_t^{(s)}(\hat{\bm{\theta}}_{j}$,$\hat{p}_{j}^{(2)})\}_{t=1}^{2n_p+1}$ and $\{\widetilde{\ell}_t^{(s)}(\hat{\bm{\theta}}_{j+1}$,$\hat{p}_{j+1}^{(2)})\}_{t=1}^{2n_p+1}$ from (\ref{PQMLeq}) based on the sample $\{\widetilde{X}_t^{(s)}\}_{t=1}^{2n_p+1}$,}\\
\multicolumn{2}{l}{~~~ compute the double-sided random walk,}\\
\end{tabular}
\end{table}
\vspace{-1cm}
\begin{align}\label{Wtau_para}
\widetilde{W}_{j,\tau}^{(s)}=\left\{ \begin{array}{lr}
\sum\limits_{t=n_p+2}^{n_p+1+\tau}\left[\widetilde{\ell}_t^{(s)}(\hat{\bm{\theta}}_{j},\hat{p}_{j}^{(2)})-\widetilde{\ell}_t^{(s)}(\hat{\bm{\theta}}_{j+1},\hat{p}_{j+1}^{(2)})\right]&\tau>0,\\
0&\tau=0,\\
\sum\limits_{t=n_p+1+\tau}^{n_p}\left[\widetilde{\ell}_t^{(s)}(\hat{\bm{\theta}}_{j+1},\hat{p}_{j+1}^{(2)})-\widetilde{\ell}_t^{(s)}(\hat{\bm{\theta}}_{j},\hat{p}_{j}^{(2)})\right]&\tau<0,
\end{array}
\right.
\end{align}
\vspace{-0.5cm}
\begin{table}[H]
\tabcolsep0.02in
		\begin{tabular}{l}
~~~ compute $\widetilde{\tau}_{j,n_p}^{(s)}=\arg\max\limits_{\tau\in\{-n_p,...,n_p\}}\widetilde{W}_{j,\tau}^{(s)}$.\\
end\\

\midrule
\emph{Obtain the final CI$_{j}^{\text{PBA}}$}:\\
\cmidrule(lr){1-1}
~~~ compute the $\frac{\alpha}{2}$ and $1-\frac{\alpha}{2}$ percentiles of the sample $\{\widetilde{\tau}_{j,n_p}^{(1)},...,\widetilde{\tau}_{j,n_p}^{(B)}\}$, denoted by $\widetilde{l}$ and $\widetilde{u}$,\\
~~~ obtain the parametric bootstrap $100(1-\alpha)\%$ CI for the change-point $\tau_j^{0}$: CI$_{j}^{\text{PBA}}=[\hat{\tau}_j^{(3)}-\widetilde{u},\hat{\tau}_j^{(3)}-\widetilde{l}].$\\
\bottomrule
\end{tabular}
\end{table}
{\bf\emph{Block Bootstrap Algorithm (BBA)}}. Unlike PBA, block bootstrap obtain replicate samples from joining block subsamples of the observations before and after the change-point.
Then similar to PBA to get the final CI$_{j}^{\text{BBA}}$. The detail of BBA for $\tau_j^{0}$ is given introduction in the following.
\begin{table}[H]
\tabcolsep0.02in
		\begin{tabular}{ll}
			\toprule
\multicolumn{2}{l}{\bf{Block Bootstrap Algorithm}:}\\
\midrule
{Input}:& The significance level $\alpha$, resample size $n_b$ and the sampling number $B$.\\
&The data set: $\{X_t\}_{t=\hat{\tau}_{j-1}^{(3)}+1}^{\hat{\tau}_{j+1}^{(3)}}$.\\
&The order estimates $\hat{p}_j^{(2)}$ and $\hat{p}_{j+1}^{(2)}$ from ``Second step".\\
&The parameter estimates $\hat{\bm{\theta}}_j$ and $\hat{\bm{\theta}}_{j+1}$, change-point estimate $\hat{\tau}_j^{(3)}$ from ``Third step".\\

\midrule
\multicolumn{2}{l}{\emph{Simulate replicate samples:}}\\
\cmidrule(lr){1-2}
\multicolumn{2}{l}{for $s=1:B$}\\
\multicolumn{2}{l}{~~~ sample the block of observations $\{X_t^{*(s)}\}_{t=1}^{n_b+1}$ from the original data set $\{X_t\}_{t=\hat{\tau}_{j-1}^{(3)}+1}^{\hat{\tau}_{j}^{(3)}}$,}\\
\multicolumn{2}{l}{~~~ sample the block of observations $\{X_t^{*(s)}\}_{t=n_b+1}^{2n_b+1}$ from the original data set $\{X_t\}_{t=\hat{\tau}_{j}^{(3)}+1}^{\hat{\tau}_{j+1}^{(3)}}$,}\\
\multicolumn{2}{l}{~~~ join them to form the resampled process $\{{X}_t^{*(s)}\}_{t=1}^{2n_b+1}$.}\\
\multicolumn{2}{l}{end}\\

\midrule
\multicolumn{2}{l}{\emph{Approximate the asymptotic distribution of} $\arg\max\limits_{\tau}W_{j,\tau}$:}\\
\cmidrule(lr){1-2}
\multicolumn{2}{l}{for $s=1:B$}\\
\multicolumn{2}{l}{~~~ obtain $\{{\ell}_t^{*(s)}(\hat{\bm{\theta}}_{j}$,$\hat{p}_{j}^{(2)})\}_{t=1}^{2n_b+1}$ and $\{\ell_t^{*(s)}(\hat{\bm{\theta}}_{j+1}$,$\hat{p}_{j+1}^{(2)})\}_{t=1}^{2n_b+1}$ from (\ref{PQMLeq}) based on the sample $\{X_t^{*(s)}\}_{t=1}^{2n_b+1}$,}\\
\multicolumn{2}{l}{~~~ compute the double-sided random walk,}\\
\end{tabular}
\end{table}
\vspace{-1cm}
\begin{align}\label{Wtau_block}
W_{j,\tau}^{*(s)}=\left\{ \begin{array}{lr}
\sum\limits_{t=n_b+2}^{n_b+1+\tau}\left[\ell_t^{*(s)}(\hat{\bm{\theta}}_{j},\hat{p}_{j}^{(2)})-\ell_t^{*(s)}(\hat{\bm{\theta}}_{j+1},\hat{p}_{j+1}^{(2)})\right]&\tau>0,\\
0&\tau=0,\\
\sum\limits_{t=n_b+1+\tau}^{n_b}\left[\ell_t^{*(s)}(\hat{\bm{\theta}}_{j+1},\hat{p}_{j+1}^{(2)})-\ell_t^{*(s)}(\hat{\bm{\theta}}_{j},\hat{p}_{j}^{(2)})\right]&\tau<0,
\end{array}
\right.
\end{align}
\vspace{-0.5cm}
\begin{table}[H]
\tabcolsep0.02in
		\begin{tabular}{l}
~~~ compute $\tau_{j,n_b}^{*(s)}=\arg\max\limits_{\tau\in\{-n_b,...,n_b\}}W_{j,\tau}^{*(s)}$.\\
end\\

\midrule
\emph{Obtain the final CI$_{j}^{\text{BBA}}$}:\\
\cmidrule(lr){1-1}
~~~ compute the $\frac{\alpha}{2}$ and $1-\frac{\alpha}{2}$ percentiles of the sample $\{\tau_{j,n_b}^{*(1)},...,\tau_{j,n_b}^{*(B)}\}$, denoted by $l^*$ and $u^*$,\\
~~~ obtain the block bootstrap $100(1-\alpha)\%$ CI for the change-point $\tau_j^{0}$: CI$_{j}^{\text{BBA}}=[\hat{\tau}_j^{(3)}-u^*,\hat{\tau}_j^{(3)}-l^*].$\\
\bottomrule
\end{tabular}
\end{table}

Next, we consider the validity of the parametric and block bootstrap procedures.
Assume that, in the parametric bootstrap procedure, the $j$th segment GCINAR model in the MCP-GCINAR model is composed of a specific thinning
operator (``$\diamond$") plus a innovation $Z_t$, that is $X_{t,j}=\sum_{k=1}^p\beta_{k,j}\diamond X_{t-k,j}+Z_{t,j}$.
And the form of the thinning operator as well as the distribution of the innovation $Z_t$ are known, then under the Assumption H.\ref{H5},
we have the following Theorem \ref{LRSM5_theory}.
For notational simplicity, we omit the superscript $(s)$ that indicates the $s$th realization of the $B$ bootstrap simulations.
\begin{assumption}\label{H5}
Denote $\mathcal{L}_{Z_{t,j},\beta}$ as the distribution of the $j$th segment innovation process $Z_{t,j}$ with the parameter $\beta\in \mathbb{R}^{+}$ ($\mathbb{R}^{+}$ is the set of positive real numbers).
Assume that ${\rm E}(Z_{t,j})^{\epsilon_Z}<\infty$ for some ${\epsilon_Z}\in \mathbb{N}$ and $0<\mathcal{L}_{Z_{t,j},\beta}(0)<1$ holds,
where $\mathcal{L}_{Z_{t,j},\beta}(k)= P(Z_{t,j}=k),k\in \mathbb{N}$.
For all $\epsilon>0$, there exist a $c>0$ such that for all $\beta'\in\mathbb{R}^{+}$ with $|\beta'-\beta|<c$, there is
$\sum_{k=0}^{\infty}|\mathcal{L}_{Z_{t,j},\beta}(k)-\mathcal{L}_{Z_{t,j},\beta'}(k)|<\epsilon$.
Furthermore, assume that there exists a neighborhood $M_{\beta^0,c}=\{\beta\big||\beta-\beta^0|<c, \beta\in\mathbb{R}^{+}\}$ of $\beta^0$
such that $\sum_{k=0}^{\infty}k^{\epsilon_Z}\mathcal{L}_{Z_{t,j},\beta}(k)<\infty$ holds uniformly on $M_{\beta^0,c}$.
\end{assumption}
\begin{theorem}\label{LRSM5_theory}
Under the setting in Theorem \ref{LRSM3_theory}, 
and for all segment, H.\ref{H5} holds, then for any fixed reasonably large $n_p\in \mathbb{N}$, $n_b\in \mathbb{N}$,
and $n_b<\min(\hat{\tau}_{j+1}^{(3)}-\hat{\tau}_{j}^{(3)},\hat{\tau}_{j}^{(3)}-\hat{\tau}_{j-1}^{(3)})$, we have
\begin{align}
&\sup\limits_{x\in \mathbb{R}}{\Big|}\widetilde{{\rm P}}(\widetilde{\tau}_{j,n_p}\leq x)-{\rm P}(\arg\max\limits_{\tau}W_{j,\tau}\leq x){\Big|}\xrightarrow{p}0,\label{theorem5.1}\\
&\sup\limits_{x\in \mathbb{R}}{\Big|}{\rm P}^{*}(\widetilde{\tau}_{j,n_b}\leq x)-{\rm P}(\arg\max\limits_{\tau}W_{j,\tau}\leq x){\Big|}\xrightarrow{p}0,\label{theorem5.2}
\end{align}
where $W_{j,\tau}$ is defined in (\ref{Wtau}), and $\widetilde{{\rm P}}$ and ${\rm P}^{*}$ denote probability measures, respectively,
under the two bootstrap schemes conditional on the original data set $\{X_t\}_{t=\hat{\tau}_{j-1}^{(3)}+1}^{\hat{\tau}_{j+1}^{(3)}}.$\\
\end{theorem}
\vspace{-5mm}

Note that a lot of model-formal assumptions, such as the form of the thinning operator and the distribution of $Z_{t,j}$, restrict the application of PBA because we usually do not have this information a priori.
To task this situation, we assume that the inference of change-points is obtained from the multiple change-points Poisson INAR (MCP-PINAR) model, where the $j$-th segment in Model (\ref{CGINAR}) is defined as follows,
\begin{align}\label{PINAR}
X_{t,j}=\sum_{k=1}^{p_j}\beta_{k,j}\bullet X_{t-k,j}+Z_{t,j},
\end{align}
where $``\bullet"$ is the Poisson thinning operator, defined as: $\beta\circ X=\sum_{i=1}^{X}\mathrm{B}_{i}(\beta)$, the counting series $\{\mathrm{B}_i(\beta)\}$ is an independent and identically distributed (i.i.d.) Poisson random sequence with mean $\beta$.
And $\{Z_{t,j}\}$ is a sequence of i.i.d. Poisson random variables with mean $\beta_{0,j}$.
Then, we do the ``simulate replicated samples" step based on the MCP-PINAR model,
and the $100(1-\alpha)\%$ CI for the change-point $\tau_j^0$ can be obtained after implementing PBA.
\begin{remark}
This setup is reasonable and convenient. On the one hand, this is a continuation of the PQML idea, which assumes that the conditional distribution is a Poisson distribution. So since there is no prior distribution information, we might as well continue with this idea.
Moreover, the subsequent simulation results also verify that the performance of confidence interval constructed by the PBA is still satisfactory when the wrong model is specified. Another advantage of this is that Theorem \ref{LRSM5_theory} can be held without assumption H.\ref{H5}, which is a complex condition after all.
\end{remark}

\section{Computational issues}
In this section, the tuning parameters and computational complexity are discussed.
\subsection{Tuning parameters setting in the LRSM }\label{Sect4.1}
The tuning parameters involved in the LRSM are $p_{\max}, m_{\max}$ and $h$.
As discussed in Section \ref{section2.1}, as long as  $p_{\max},m_{\max}$ are large enough,
their values have little effect on the estimation results.
In contrast, the LRSM estimates are mainly affected by the window parameter $h$.
Smaller windows are more sensitive to changes  occurring in short time durations,
and are more likely to guarantee the condition, $2h<n\epsilon_{\tau}$,
while larger windows are more sensitive to small changes.
It is theoretically crucial to choose an $h>d(\log n)^3$ for Theorem \ref{LRSM1_theory} to hold, where $d$ is an unknown window coefficient constant.
Based on the experience of scholars as well as our simulation studies, setting $h=\max(n/20,(\log n)^4/25)$ will generally yield satisfactory results with different models and sample sizes.

However, such a setting may not guarantee the condition $2h<n\epsilon_{\tau}$ hold, when the sample size is small and the change-points are dense.
To solve this, we can combine multiple $h$ in the first step of LRSM to get potential change-points.
The detailed operation is explained as follows,

Setting window coefficient set $\tilde{d}$ (such as $\tilde{d}=\{1,2\}$), and $h_{mix}=\{h_i\}_{i=1}^{|\tilde{d}|}$, where $h_i=\tilde{d}_i(\log n)^4/25$ (such as $h_1=(\log n)^4/25, h_2=2(\log n)^4/25)$. Then start from $i=1$ to $i=|\tilde{d}|$, get the potential change-points set $\hat{\mathcal{J}}^{(1)}(h_i)$, combine them and get the final $\hat{\mathcal{J}}^{(1)}(h)$.

Similar aggregation procedures can be found in \cite{Ng2022}.
Empirical evidence suggests that, setting $\tilde{d}=\{0.2,0.4,0.6,0.8,1,1.2\}$ for small sample size and $\tilde{d}=\{1,2,3,4,5,6\}$ for large sample size usually results in more satisfactory performance.

\subsection{Tuning parameters setting in Bootstrap procedure}
The tuning parameters involved in the two bootstrap methods are the sampling number $B$, resample size $n_p$ and $n_b$.
For the sampling number $B$, large enough to guarantee reliable results, such as $B=1000$.
For the parametric bootstrap, we set $n_p=n/2$ to ensure that $n_p$ is sufficiently large to include the maxima of the random walk.
For the block bootstrap, the following iterative procedure is used to find a data-driven or adaptive block bandwidth $n_b$:

$\bullet$ \textbf{Step 1:} Compute a $(1-\alpha)\%$ CI using CIs' approximation or the PBA method. Denote
the width of the CI by $l$. Set initial block bandwidth $n_b = 2l$.

$\bullet$ \textbf{Step 2:} Perform the BBA to obtain $\{\tau_{j,n_b}^{*(s)}\}, s= 1,2,...,B.$
If the proportion of the sample $\{\tau_{j,n_b}^{*(s)}\}$ lying in either of the regions $[\hat{\tau}_{j}^{(3)}-n_b, \hat{\tau}_{j}^{(3)}-(1-\alpha)n_b]$ or $[\hat{\tau}_{j}^{(3)}+(1-\alpha)n_b, \hat{\tau}_{j}^{(3)}+n_b]$ is greater than $(\alpha/2)\%$, increase $n_b$ by $l$.

$\bullet$ \textbf{Step 3:} Repeat \textbf{Step 2:} until a final block bandwidth $n_b$ is found. Set this $n_b$ as the adaptive bandwidth.
\subsection{Computational complexity}\label{sect4.4}
The computational complexity of the three-step LRSM has been discussed by many scholars, such as \cite{YauZhao2016}, \cite{Ng2022}.
Since the size of $W_t(h)$ is $2h$, the computational complexity of $S_h(t)$ for each $t$ is order $O(h)$, and the computational complexity of the first step in LRSM is $O(nh)$.
The computational complexity of the second step, the OP Algorithm, is $O\big((\hat{m}^{(1)})^2 n\big)$ and the computational complexity of the third step for evaluating extended local windows is $O\big(\hat{m}^{(2)}h^2 \big)$.
Therefore, the total computational complexity in the three-step LRSM procedure is $O(hn).$
\section{Simulation}\label{Set5}
To evaluate the finite-sample performance of the proposed LRSM and Bootstrap procedures, we conducted extensive simulation studies and split the simulation studies into the following five parts.
In the first part, we consider the sensitivity of the LRSM to its tuning parameter, the windows parameter $h$.
In the second simulation,
we compare the performance of LRSM with GA and penQLIK.
Furthermore, the performance of LRSM to deal with a large number of change-points under different sample size is compared in the third part.
Finally, the performance of the CIs constructed by the parametric and block bootstrap approximations proposed in Section \ref{sect2.5} is examined, and compared to the approximation method of CIs, proposed in Theorem \ref{asym}.

We first introduce the following notations and evaluation metrics that are primarily used in simulations.\\
\vspace{-5mm}~\\
{\bf\emph{Notations:}}

$\bullet$ Denote $``\circ"$ as the binomial thinning operator, which is proposed by \cite{{SV1979}} and defined as: $\beta\circ X=\sum_{i=1}^{X}\mathrm{B}_{i}(\beta)$, the counting series $\{\mathrm{B}_i(\beta)\}$ is an independent and identically distributed (i.i.d.) Bernoulli random sequence with mean $\beta$.

$\bullet$ Denote $``\ast$" as the negative binomial thinning operator of \cite{Ristic2009}, it is defined as $\beta\ast X=\sum_{i=1}^X \mathrm{B}_i\big(\beta/(1+\beta)\big)$, the counting series $\{\mathrm{B}_i\big(\beta/(1+\beta)\big)\}$ is a sequence of i.i.d. geometric random variables with mean $\beta$,

$\bullet$ Denote the Poisson distribution with mean $\beta$ by ${\rm Poi}(\beta)$,
the Geometric distribution with mean $\beta$ by ${\rm Geo}(\beta/(1+\beta))$.

$\bullet$  Let ${\bm\upsilon}_n=(\upsilon_1,...,\upsilon_{m})$, $0<\upsilon_1<...<\upsilon_{m}<1$, satisfy $\tau_j=[\upsilon_jn]$, where $[x]$ is the greatest integer that is less than or equal to $x$.\\
\vspace{-5mm}~\\
{\bf\emph{Evaluation Metrics:}}

$\bullet$ Denote the true positive rate of $m$ by TPR($m$), i.e., the proportion of informative points are correctly identified.

$\bullet$ Define the following two type evaluation metrics to measure the under-segmentation error and the over-segmentation error of the change-points location estimate $\bm{\hat{\upsilon}_n}$, respectively.
\begin{align*}
\zeta_{u}(\bm{\upsilon}^0|\hat{\bm{\upsilon}}_n)=\sup\limits_{b\in\hat{\bm{\upsilon}}_n}\inf\limits_{a\in\bm{\upsilon}^0}|a-b|,
~~~\zeta_{o}(\hat{\bm{\upsilon}}_n|\bm{\upsilon}^0)=\sup\limits_{b\in\bm{\upsilon}^0}\inf\limits_{a\in\hat{\bm{\upsilon}}_n}|a-b|~\text{\citep{Boysen2009}}.
\end{align*}
A desirable estimate should be able to balance both quantities.

$\bullet$ Define the following metric to measure the location accuracy of estimated change-points.
$$\zeta_{d}(\hat{\bm{\upsilon}}_n,\bm{\upsilon}^0)=\frac{1}{|\bm{\upsilon}^0|}\sum\limits_{\upsilon_k^0\in\bm{\upsilon}^0}\min\limits_{\hat{\upsilon}_j\in\hat{\bm{\upsilon}}_n}|\hat{\upsilon}_j-\upsilon_k^0| ~~~\text{\citep{Chen2021}},$$
which is the distance from the estimated set $\hat{\bm{\upsilon}}_n$ and the true change-points set $\bm{\upsilon}^0$. 

All simulations are carried out using the MATLAB software. The empirical results displayed in the tables are computed over $1000$ replications.
\subsection{Sensitivity analysis for the tuning parameter $h$ in the LRSM}
To study the choice of the tuning parameter $h$ in the three-step LRSM, we conducted a sensitivity analysis by using data generated from
Model (A1) with the sample sizes $n=500, 1000, 2000$.
$h=\max(n/20,d(\log n)^4/25)$ was considered using different values of $d$.
Furthermore, the combine multiple $h$, introduced in Section \ref{Sect4.1}, was also studied with $d_{mix}=\{0.2,0.4,0.6,0.8,1,1.2\}$.
The results are summarized in Table \ref{sensitive_h}.\\
Model (A1) is as follows:
\begin{align*}
X_t=\left\{ \begin{array}{lll}
0.5\circ X_{t-1,1}+Z_t&Z_t\overset{i.i.d}{\sim} {\rm Poi}(0.5)&0<t\leq\tau_1^0,\\
0.126\circ X_{t-\tau_1^0-1,2}+0.254\circ X_{t-\tau_1^0-2,2}+0.297\circ X_{t-\tau_1^0-3,2}+Z_t&Z_t\overset{i.i.d}{\sim} {\rm Poi}(1)&\tau_1^0<t\leq\tau_2^0,\\
0.4\circ X_{t-\tau_2^0-1,3}+Z_t&Z_t\overset{i.i.d}{\sim} {\rm Poi}(2)&\tau_2^0<t\leq n.
\end{array}\right.
\end{align*}
where $(\tau_1^0,\tau_2^0)=([0.3n],[0.6n])$.

\begin{table}
\scriptsize
\tabcolsep0.02in
	\centering
	\caption{Sensitivity analysis for the window $h$ in the LRSM based on Model (A1).}\label{sensitive_h}
	\begin{tabular}{*{8}{c}}
\toprule
$d$&$h=\max(\frac{n}{20},d\frac{(\log n)^4}{25})$&Sample size ($n$)&TRP($m$)&$\zeta_{u}(\bm{\upsilon}^0|\hat{\bm{\upsilon}}_n)$&$\zeta_{o}(\hat{\bm{\upsilon}}_n|\bm{\upsilon}^0)$&$\zeta_{d}(\hat{\bm{\upsilon}}_n,\bm{\upsilon}^0)$&Average time ($s$)\\\cmidrule(lr){1-3}\cmidrule(lr){4-8}
0.5&29&500&0.8070 &0.0410 &0.0973 &0.1645 &0.7439 \\
0.5&50&1000&0.9930 &0.0284 &0.0287 &0.0484 &1.5489 \\
0.5&100&2000&0.9980 &0.0107 &0.0105 &0.0178 &3.6764 \vspace{1mm}\\
1&59&500&0.7110 &0.0370 &0.1219 &0.2059 &0.5506 \\
1&91&1000&0.9760 &0.0250 &0.0316 &0.0518 &1.5935 \\
1&133&2000&0.9890 &0.0110 &0.0102 &0.0172 &4.3840 \vspace{1mm}\\
1.5&89&500&0.6710 &0.0274 &0.1237 &0.2071 &0.7409 \\
1.5&136&1000&0.9690 &0.0163 &0.0248 &0.0414 &2.2516 \\
1.5&200&2000&0.9950 &0.0063 &0.0060 &0.0108 &8.5836 \vspace{1mm}\\
2&119&500&0.6400 &0.0234 &0.1290 &0.2111 &1.6022 \\
2&182&1000&0.9580 &0.0113 &0.0238 &0.0401 &4.9702 \\
2&267&2000&0.9990 &0.0055 &0.0054 &0.0100 &13.1729 \vspace{1mm}\\
2.5&149&500&0.3660 &0.0383 &0.2214 &0.3714 &1.6629 \\
2.5&227&1000&0.9030 &0.0118 &0.0398 &0.0647 &5.8175 \\
2.5&333&2000&0.9900 &0.0054 &0.0078 &0.0135 &14.9413 \vspace{1mm}\\
3&178&500&0.0000 &0.0892 &0.3831 &0.7040 &0.7435 \\
3&273&1000&0.8010 &0.0132 &0.0716 &0.1127 &3.8218 \\
3&400&2000&0.9880 &0.0053 &0.0088 &0.0152 &11.0657 \\\midrule
$d_{mix}$&$h_{mix}=d_{mix}\frac{(\log n)^4}{25}$&Sample size ($n$)&TRP($m$)&$\zeta_{u}(\bm{\upsilon}^0|\hat{\bm{\upsilon}}_n)$&$\zeta_{o}(\hat{\bm{\upsilon}}_n|\bm{\upsilon}^0)$&$\zeta_{d}(\hat{\bm{\upsilon}}_n,\bm{\upsilon}^0)$&Average time ($s$)\\\cmidrule(lr){1-3}\cmidrule(lr){4-8}
$\{0.2,0.4,0.6,0.8,1,1.2\}$&$\{12,24,36,48,59,71\}$&500&0.9130 &0.0276 &0.0529 &0.0909 &3.7002 \\
$\{0.2,0.4,0.6,0.8,1,1.2\}$&$\{19,37,55,73,91,110\}$&1000&1.0000 &0.0166 &0.0166 &0.0301 &9.9580 \\
$\{0.2,0.4,0.6,0.8,1,1.2\}$&$\{27,54,80,107,133,160\}$&2000&0.9990 &0.0099 &0.0096 &0.0166 &31.0050 \\
\bottomrule
	\end{tabular}
\end{table}
The following conclusions can be drawn from Table \ref{sensitive_h}.
In terms of the calculation time (Average time) and the metric TRP, smaller windows reduce computing cost, and are more sensitive to changes occurring in short time durations.
In terms of the metrics $\zeta_{u}(\bm{\upsilon}^0|\hat{\bm{\upsilon}}_n)$, $\zeta_{o}(\hat{\bm{\upsilon}}_n|\bm{\upsilon}^0)$, and $\zeta_{d}(\hat{\bm{\upsilon}}_n,\bm{\upsilon}^0)$, larger windows balance the under-segmentation error
and the over-segmentation error, improving the accuracy of estimates.
Such as, $n=2000$, $h=200, 267$ have the optimal accuracy $0.0108$ and $0.01$ in the entire result table.
In particular, the mixture window $h_{mix}$ has excellent performance, of course, this advantage is at the cost of increasing the calculation cost.
In addition, the LRSM is robust given a mild violation of $2h<n\epsilon_{\tau}$.
Such as, with the sample size $n=1000, \tau_1^0=300$ and $\tau_2^0=600$, the minimum distance being $300$,  the LRSM requires $h<300/2=150$ for consistent estimate. However, the estimate remains satisfactory when $h=273$.
This may be related to the findings that the likelihood ratio test statistic can consistently estimate
a single change-point in the presence of multiple changes.
Thus, although the choice of $h$ will affect the accuracy of estimation results, it is not sensitive for estimation purposes.
\subsection{Comparing LRSM and existing methods}
In this subsection, we compare the performance of GA \citep{Sheng2023}, penalized contrast QLIK (penQLIK, \cite{DiopKengne2021b}),
and the LRSM in implementing change-point inference. We consider Model (A1) with sample sizes of $n=500, 1000, 2000$.
Additionally, we include the penQLIK method proposed by \cite{DiopKengne2021b} for comparison with the LRSM.
The penQLIK is a data-driven method that utilizes the slope heuristic procedure \citep{Baudry2012} to calibrate the penalty term. The penQLIK criterion is minimized using a dynamic programming algorithm. It is important to note that penQLIK does not consider the change of orders $p$. Therefore, we implement penQLIK by considering a sufficiently large $p=5,~q=5$ (denoted as penQLIK(5,5)).
\begin{table}[H]
\tabcolsep0.08in
	\centering
	\caption{Comparison results of the penQLIK, GA and LRSM based on Model (A1).}\label{compare}
	\begin{tabular}{*{7}{c}}
\toprule
Method&Sample size($n$)&TRP($m$)&$\zeta_{u}(\bm{\upsilon}^0|\hat{\bm{\upsilon}}_n)$&$\zeta_{o}(\hat{\bm{\upsilon}}_n|\bm{\upsilon}^0)$&$\zeta_{d}(\hat{\bm{\upsilon}}_n,\bm{\upsilon}^0)$&Average time ($s$)\\\cmidrule(lr){1-2}\cmidrule(lr){1-2}\cmidrule(lr){3-7}
penQLIK(5,5)&500&0.4190 &0.1250 &0.0440 &0.0772 &129.9832 \\
&1000&0.7140 &0.0916 &0.0172 &0.0298 &935.9727 \\
&2000&0.8290 &0.1087 &0.0073 &0.0127 &5697.9496\\\midrule
GA&500&0.9090 &0.0174 &0.0437 &0.0721 &1052.2157 \\
&1000&0.9990 &0.0087 &0.0090 &0.0162 &3217.8543 \\
&2000&1.0000 &0.0043 &0.0043 &0.0078 &3688.3210\\\midrule
LRSM&500&0.8070 &0.0410 &0.0973 &0.1645 &0.7439 \\
&1000&0.9930 &0.0284 &0.0287 &0.0484 &1.5489 \\
&2000&0.9980 &0.0107 &0.0105 &0.0178 &3.6764 \\
\bottomrule
	\end{tabular}
\end{table}
As you can see from Table \ref{compare}, GA is the top performer in almost all evaluation metrics.
It has a smaller $\zeta_{d}(\widehat{\widetilde{\bm{\tau}}},\widetilde{\bm{\tau}}^0)$, the highest TRP($m$), and balances $\zeta_{u}(\widetilde{\bm{\tau}}^0|\widehat{\widetilde{\bm{\tau}}})$ and $\zeta_{o}(\widehat{\widetilde{\bm{\tau}}}|\widetilde{\bm{\tau}}^0)$.
However, the results of the LRSM are also acceptable, especially in Model (A1) with sample size $n=2000$.
It balances $\zeta_{u}(\widetilde{\bm{\tau}}^0|\widehat{\widetilde{\bm{\tau}}})$ and $\zeta_{o}(\widehat{\widetilde{\bm{\tau}}}|\widetilde{\bm{\tau}}^0)$ and the value of TRP($m$) is closed to 1.
In addition, although penQLIK can handle this sequence-changing model, it produces a large number of under-segmentation errors. In other words, penQLIK does not compress the model any better than the MDL criterion.
Taking into account all simulation results, LRSM has good and stable performance.
Especially considering the calculation time, the advantage is more obvious.
In summary, GA achieves the global optimization of MDL, so the optimal results are given in terms of estimation accuracy.
However, when the sample size is large and the number of change-points is relatively small, GA may require computational cost, and the LRSM can serve as a reliable and computationally efficient alternative.

\subsection{Performance of the LRSM based on difference scenarios}
In this {\color{red}subection}, the performance of the LRSM is studied based on the scenario of a large number of dense change-points with a finite-sample (sample size $n=2000$) and in the scenario of a long time series (sample size $n=10000$).
Two types of the MCP-GCINAR model are considered, B-type models have a Poisson distribution innovation and a binomial thinning operator, while C-type models have a negative binomial thinning operator and a Geometric distribution innovation.
Models (B1) - (B9) represent the case of B-type models under $m_0=1$ to $m_0=9$,
and Models (C1) - (C9) represent the case of C-type models under change-points numbers $m_0=1$ to $m_0=9$.

While sample size $n=2000$, the LRSM is implemented by the $h=(\log n)^4/25=133$ and $h_{mix}=\{27,54,80,107,133,160\}$,
and while in the scenario of a long-time series, sample size $n=10000$, the LRSM is implemented by the $h=(\log n)^4/25=287$.
All results of Models (B1) - (B9)  are summarized in Table \ref{change_m_2000_bin} and Table \ref{change_m_10000_bin}.
The specific forms of Models (B1) - (C9) are in the Appendix to save space.

Table \ref{change_m_2000_bin} reveals that the application of $h_{mix}$ can improve LRSM's performance, because it significantly improves the metric TRP, reduces the metrics $\zeta_{u}(\bm{\upsilon}^0|\hat{\bm{\upsilon}}_n)$, $\zeta_{o}(\hat{\bm{\upsilon}}_n|\bm{\upsilon}^0)$, and $\zeta_{d}(\hat{\bm{\upsilon}}_n,\bm{\upsilon}^0)$, and balances the under-segmentation error and the over-segmentation error.
But this apparent advantage also comes at the cost of increasing the computational burden.
Therefore, in the scenario of large sample size ($n=10000$), Table \ref{change_m_10000_bin} only summarizes the results of applying LRSM with $h=287$.
After all, considering  $h_{mix}$, a lot of time and computing resources are needed.
However, it can be seen that in such a situation, the general $h$ can handle the change-point problem of the MCP-GCINAR models well.
\begin{adjustwidth}{-1cm}{-1cm}
\begin{table}
\scriptsize
\tabcolsep0.015in
	\centering
	\caption{Performance of the three-step LRSM based on Models (B1) - (C9) with sample size $n=2000$, $h=133$ and  $h_{mix}=\{27,54,80,107,133,160\}$.}\label{change_m_2000_bin}
	\begin{tabular}{*{16}{c}}
\toprule
&&&&&&&Average&&&&&&&&Average\\
Model&$m_0$&window&TRP($m$)&$\zeta_{u}(\bm{\upsilon}^0|\hat{\bm{\upsilon}}_n)$&$\zeta_{o}(\hat{\bm{\upsilon}}_n|\bm{\upsilon}^0)$&$\zeta_{d}(\hat{\bm{\upsilon}}_n,\bm{\upsilon}^0)$&time ($s$)&Model&$m_0$&window&TRP($m$)&$\zeta_{u}(\bm{\upsilon}^0|\hat{\bm{\upsilon}}_n)$&$\zeta_{o}(\hat{\bm{\upsilon}}_n|\bm{\upsilon}^0)$&$\zeta_{d}(\hat{\bm{\upsilon}}_n,\bm{\upsilon}^0)$&time ($s$)\\\cmidrule(lr){1-3}\cmidrule(lr){4-8}\cmidrule(lr){9-11}\cmidrule(lr){12-16}
(B1)&1&$h$&0.9890 &0.0034 &0.0026 &0.0052 &6.9152 &(C1)&1&$h$&0.9590 &0.0085 &0.0037 &0.0073 &6.9470 \\
&&$h_{mix}$&0.9990 &0.0026 &0.0023 &0.0047 &49.3130 &&&$h_{mix}$&0.9780 &0.0078 &0.0031 &0.0062 &50.2625 \\
(B2)&2&$h$&0.9860 &0.0066 &0.0050 &0.0084 &7.6469 &(C2)&2&$h$&0.9540 &0.0124 &0.0062 &0.0108 &7.6078 \\
&&$h_{mix}$&0.9980 &0.0028 &0.0027 &0.0049 &53.1847 &&&$h_{mix}$&0.9710 &0.0084 &0.0044 &0.0081 &51.9576 \\
(B3)&3&$h$&0.9860 &0.0052 &0.0065 &0.0089 &8.0786 &(C3)&3&$h$&0.9700 &0.0075 &0.0064 &0.0095 &8.0287 \\
&&$h_{mix}$&1.0000 &0.0034 &0.0034 &0.0056 &43.0814 &&&$h_{mix}$&0.9800 &0.0064 &0.0048 &0.0074 &43.2224 \\
(B4)&4&$h$&0.9740 &0.0048 &0.0099 &0.0127 &8.8324 &(C4)&4&$h$&0.9710 &0.0073 &0.0079 &0.0109 &8.8331 \\
&&$h_{mix}$&0.9970 &0.0041 &0.0040 &0.0066 &46.3998 &&&$h_{mix}$&0.9560 &0.0085 &0.0055 &0.0085 &46.1871 \\
(B5)&5&$h$&0.7260 &0.0073 &0.0687 &0.0683 &9.3995 &(C5)&5&$h$&0.9500 &0.0086 &0.0136 &0.0152 &9.4700 \\
&&$h_{mix}$&0.9580 &0.0070 &0.0135 &0.0150 &48.0895 &&&$h_{mix}$&0.8810 &0.0135 &0.0074 &0.0101 &48.1504 \\
(B6)&6&$h$&0.0110 &0.0113 &0.1640 &0.2514 &8.7535 &(C6)&6&$h$&0.0860 &0.0144 &0.1127 &0.1484 &9.0956 \\
&&$h_{mix}$&0.3690 &0.0193 &0.0868 &0.0868 &41.7372 &&&$h_{mix}$&0.8960 &0.0229 &0.0463 &0.0497 &42.9014 \\
(B7)&7&$h$&0.0430 &0.0180 &0.4030 &0.8972 &8.0163 &(C7)&7&$h$&0.2140 &0.0353 &0.1097 &0.1374 &10.3091 \\
&&$h_{mix}$&0.7630 &0.0254 &0.0604 &0.0857 &41.4308 &&&$h_{mix}$&0.6180 &0.0456 &0.0384 &0.0436 &45.9520 \\
(B8)&8&$h$&0.0090 &0.0258 &0.6544 &1.6095 &7.3036 &(C8)&8&$h$&0.1220 &0.0231 &0.1261 &0.1748 &10.6836 \\
&&$h_{mix}$&0.6390 &0.0264 &0.1274 &0.2370 &42.7765 &&&$h_{mix}$&0.7900 &0.0300 &0.0407 &0.0442 &45.5573 \\
(B9)&9&$h$&0.0010 &0.0147 &0.4927 &1.2206 &7.7363 &(C9)&9&$h$&0.0310 &0.0196 &0.1323 &0.2141 &7.0678 \\
&&$h_{mix}$&0.2180 &0.0216 &0.2059 &0.3833 &46.0448 &&&$h_{mix}$&0.8770 &0.0298 &0.0528 &0.0550 &42.8664 \\

\bottomrule
	\end{tabular}
\end{table}
\end{adjustwidth}
\begin{table}
\scriptsize
\tabcolsep0.015in
	\centering
	\caption{Performance of the three-step LRSM based on Models (B1) - (C9) with sample size $n=10000$ and $h=287$.}\label{change_m_10000_bin}
	\begin{tabular}{*{14}{c}}
\toprule
Model&$m_0$&TRP($m$)&$\zeta_{u}(\bm{\upsilon}^0|\hat{\bm{\upsilon}}_n)$&$\zeta_{o}(\hat{\bm{\upsilon}}_n|\bm{\upsilon}^0)$&$\zeta_{d}(\hat{\bm{\upsilon}}_n,\bm{\upsilon}^0)$&Average time ($s$)&Model&$m_0$&TRP($m$)&$\zeta_{u}(\bm{\upsilon}^0|\hat{\bm{\upsilon}}_n)$&$\zeta_{o}(\hat{\bm{\upsilon}}_n|\bm{\upsilon}^0)$&$\zeta_{d}(\hat{\bm{\upsilon}}_n,\bm{\upsilon}^0)$&Average time ($s$)\\\cmidrule(lr){1-2}\cmidrule(lr){3-7}\cmidrule(lr){8-9}\cmidrule(lr){10-14}
(B1)&1&0.99 &0.0007 &0.0005 &0.0010 &69.4290 &(C1)&1&1.00 &0.0008 &0.0008 &0.0015 &69.01765\\
(B2)&2&0.94 &0.0040 &0.0006 &0.0010 &70.5583 &(C2)&2&0.93 &0.0046 &0.0009 &0.0016 &71.3397\\
(B3)&3&0.98 &0.0012 &0.0008 &0.0013 &69.3625 &(C3)&3&0.89 &0.0081 &0.0013 &0.0019 &70.1191\\
(B4)&4&0.97 &0.0031 &0.0008 &0.0014 &73.3520 &(C4)&4&0.84 &0.0111 &0.0010 &0.0016 &73.67172\\
(B5)&5&0.96 &0.0020 &0.0010 &0.0015 &75.4675 &(C5)&5&0.87 &0.0096 &0.0014 &0.0019 &76.34772\\
(B6)&6&0.82 &0.0026 &0.0210 &0.0169 &74.3734 &(C6)&6&0.92 &0.0065 &0.0030 &0.0036 &75.81408\\
(B7)&7&0.99 &0.0149 &0.0130 &0.0105 &78.2058 &(C7)&7&0.73 &0.0268 &0.0033 &0.0037 &80.13946\\
(B8)&8&0.94 &0.0110 &0.0113 &0.0087 &79.4421 &(C8)&8&0.94 &0.0054 &0.0026 &0.0031 &81.14015\\
(B9)&9&0.80 &0.0041 &0.0239 &0.0166 &81.5739 &(C9)&9&0.98 &0.0034 &0.0026 &0.0032 &81.07215\\
\bottomrule
	\end{tabular}
\end{table}
%

\subsection{Constructing confidence intervals}
In this subsection, we examine the coverage accuracy of the CIs for the change-points estimates based on the correct estimates of Models (B1) - (C2) in the previous subsection.
For each of the cases, CIs using CIs approximation (\ref{CI_appromix}),
parametric bootstrap and block bootstrap with adaptive block bandwidth are generated.

Table \ref{lrsm_ci} provides a summary of the CIs results based on the LRSM procedure, including the median and mean of the estimate $\hat{\tau}_j^{(3)}$,
the range of the middle 90\% of the final estimates (90\% Range)
average of the end-points (Mean of 90\% CI), and coverage probability (Coverage Prob.) of the CIs.

From the results in Table \ref{lrsm_ci}, it can be seen that the CIs of both PBA and BBA methods have quite accurate coverage probability,
which are close to the nominal level of 90\%.
Particularly, the CIs constructed from the two bootstrap procedures have more accurate coverage
probability than that obtained from the approximation method (\ref{CI_appromix}).
This is consistent with the previous discussion,
the pivotal approximations in CI (\ref{CI}) work unsatisfactorily under medium and large parameter changes.\vspace{-1mm}
\begin{table}
\scriptsize
\tabcolsep0.05in
	\centering
	\caption{Confidence intervals based on the LRSM change-points correct estimates\\ under the Models (B1) to (C2).}\label{lrsm_ci}
	\begin{tabular}{*{4}{c}l*{3}{c}}
\toprule
&&Median of&Mean of&\multicolumn{1}{c}{Construct CIs}&90\%&Mean of&Coverage\\
Model&$\tau_j^0$&$\hat{\tau}_j^{(3)}$&$\hat{\tau}_j^{(3)}$&\multicolumn{1}{c}{Method}&range&90\% CI&Prob.\\\cmidrule(lr){1-8}
Model (B1)&1000&998&996.2961 &CIs' approximation&[993,1003]&$[990.83 ,1001.76 ]$&74.32 \%\\
&1000&&&Parametric bootstrap&[956,1006]&$[952.64 ,1004.14 ]$&89.02 \%\\
&1000&&&Block bootstrap($n_b=50$)&[971,1005]&$[968.61 ,1002.91 ]$&85.50 \%\vspace{1mm}\\
Model (C1)&1000&999&999.8216 &CIs' approximation&[993,1006]&$[993.17 ,1006.48 ]$&70.85 \%\\
&1000&&&Parametric bootstrap&[959,1007]&$[957.22 ,1007.80 ]$&89.50 \%\\
&1000&&&Block bootstrap($n_b=50$)&[965,1006]&$[964.98 ,1007.02 ]$&86.65 \%\vspace{1mm}\\
Model (B2)&600&598&596.1302 &CIs' approximation&[593,604]&$[590.68 ,601.58 ]$&71.44 \%\\
&600&&&Parametric bootstrap&[557,606]&$[552.56 ,604.24 ]$&90.01 \%\\
&600&&&Block bootstrap($n_b=42$)&[572,605]&$[570.06 ,603.07 ]$&85.07 \%\\
&1200&1200&1203.3673 &CIs' approximation&[1196,1203]&$[1199.80 ,1206.94 ]$&87.08 \%\\
&1200&&&Parametric bootstrap&[1193,1204]&$[1196.12 ,1207.71 ]$&92.43 \%\\
&1200&&&Block bootstrap($n_b=18$)&[1195,1204]&$[1198.34 ,1207.50 ]$&90.21 \%\vspace{1mm}\\
Model (C2)&600&599&598.7670 &CIs' approximation&[592,606]&$[591.94 ,605.59 ]$&68.45 \%\\
&600&&&Parametric bootstrap&[559,607]&$[556.34 ,606.83 ]$&87.63 \%\\
&600&&&Block bootstrap($n_b=42$)&[567,606]&$[567.43 ,606.36 ]$&84.64 \%\\
&1200&1200&1200.9876 &CIs' approximation&[1196,1204]&$[1196.65 ,1205.33 ]$&76.80 \%\\
&1200&&&Parametric bootstrap&[1193,1204]&$[1193.90 ,1205.53 ]$&81.03 \%\\
&1200&&&Block bootstrap($n_b=18$)&[1192,1205]&$[1192.98 ,1206.61 ]$&83.20 \%\\
\bottomrule
	\end{tabular}
\end{table}

\section{Real data}\label{Set6}
In this section, 
the LRSM was used to detect the change-points in the daily stock volume data of four airline groups.
The aim is to investigate whether any of the detected change-points were influenced by recent sudden international events, particularly the global COVID-19 over the past few years.
Four airline data sets were originally downloaded it on-line at the Yahoo Finance web site (\url{https://hk.finance.yahoo.com/}), including:\\
$\bullet$ Singapore Airlines Limited (C6L.SI): the series has 5,927 observations from January 4, 2000, to May 19, 2023.\\
$\bullet$ Air China Limited (0753.HK): the series has 4,550 observations from December 16, 2004, to May 19, 2023.\\
$\bullet$ Deutsche Lufthansa AG (LHA.DE): the series has 6,774 observations from December 17, 1996, to May 19, 2023.\\
$\bullet$ United Airlines Holdings, Inc. (UAL): the series has 4,338 observations from February 27, 2006, to May 19, 2023.

\begin{figure}[H]
\begin{adjustwidth}{-2cm}{-1cm}
\begin{center}
\includegraphics[width=8in,height=4in]{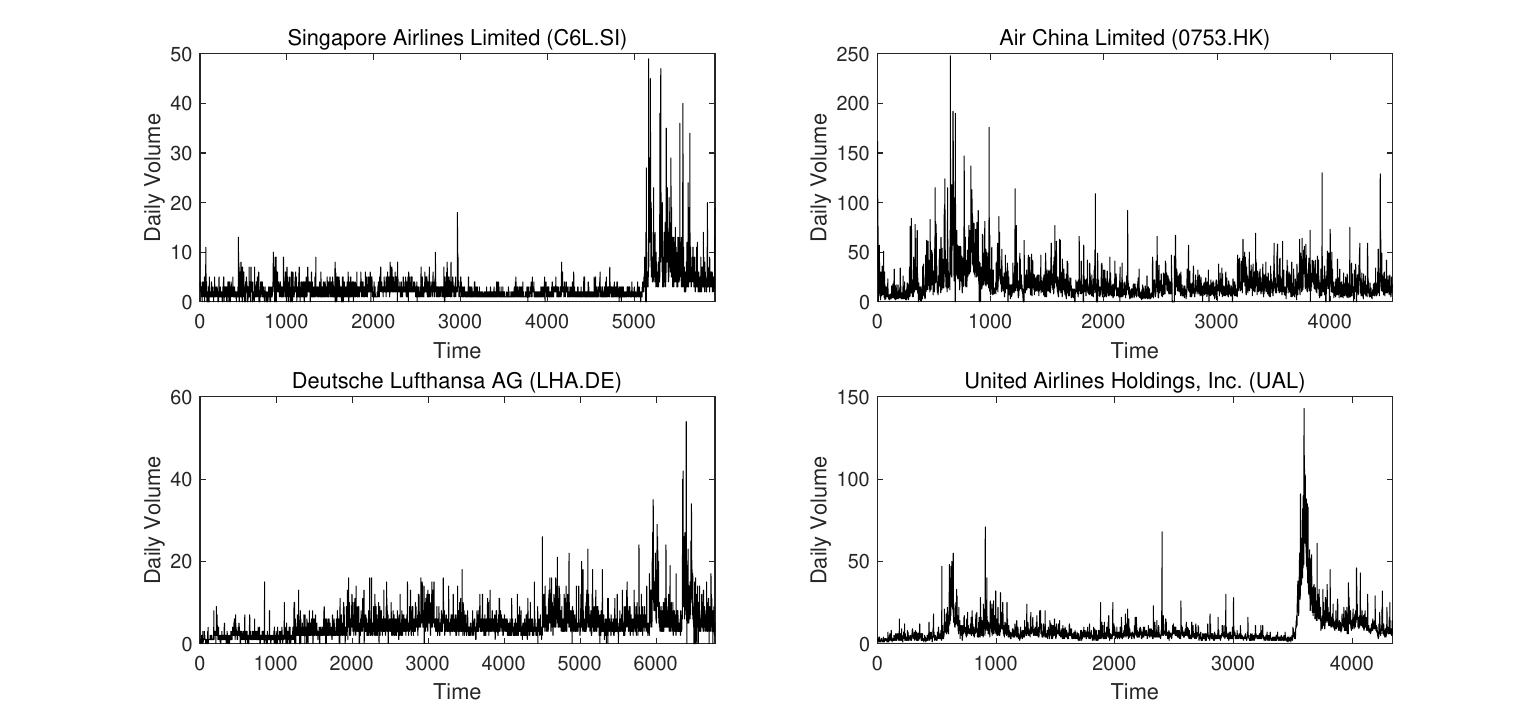}
\end{center}
\end{adjustwidth}
\vspace{-5mm}
\caption{Sample paths for six airline group daily trading volume data sets,\\including C6L.SI, 0753.HK, LHA.DE, UAL.}
\label{Sample-plot}
\end{figure}

As the data is relatively large, considering the convenience of calculation, we analyze the data by the unit of `1e+06' trading volume.
Figure \ref{Sample-plot} shows sample path figures for six data sets. It can be roughly judged from the figure that there are obvious change-points in the company's trading volume data.
We applied the three-step LRSM with $h=(\log n)^4/25$, $m_{max}=30$, $p_{max}=7$ to implement the change-points inference.
Furthermore, we evaluated the goodness of fit of the model using root mean square of differences between observations and forecasts (RMS) and Pearson residuals (Pr). These evaluation metrics are commonly employed. The equations defining RMS and Pr are provided below.
\begin{align*}
{\rm RMS}=\sqrt{\frac{1}{n-1}\sum\limits_{t=2}^{n}\left(X_t-{\rm E}(X_t|\mathcal{F}_{t-1})\right)^2},~~~~~
{\rm Pr}_t
=\frac{X_t-{\rm E}(X_t|\mathcal{F}_{t-1})}{\sqrt{{\rm Var}(X_t|\mathcal{F}_{t-1})}}.
\end{align*}
Table \ref{real_data} summarizes all change-points inference results for four airline group data sets,
including the estimate of the number of change-points ($\hat{m}$),
the estimate of change-points locations ($\hat{\tau}$) and their corresponding time (Date),
the MDL value, RMS and the mean and variance of ${\rm Pr}_t$ (Pr$_{\rm m}$ and Pr$_{\rm Var}$).
In addition, two bootstrap methods, PBA and BBA, are applied to obtain CIs of change-points estimates, where the adaptive block bandwidth $n_b$ in BBA is in parentheses.
In order to be intuitive, the change-points of each model and the CIs constructed by two bootstrap methods are shown in the figures.
However, due to space constraints, only one group (C6L.SI) is listed (Model \ref{estimate_C6L} and Figure \ref{estimate_C6L_f}); the others are in the Appendix.

\begin{figure}
\begin{adjustwidth}{-1cm}{-1cm}
\includegraphics[width=7.5in,height=2in]{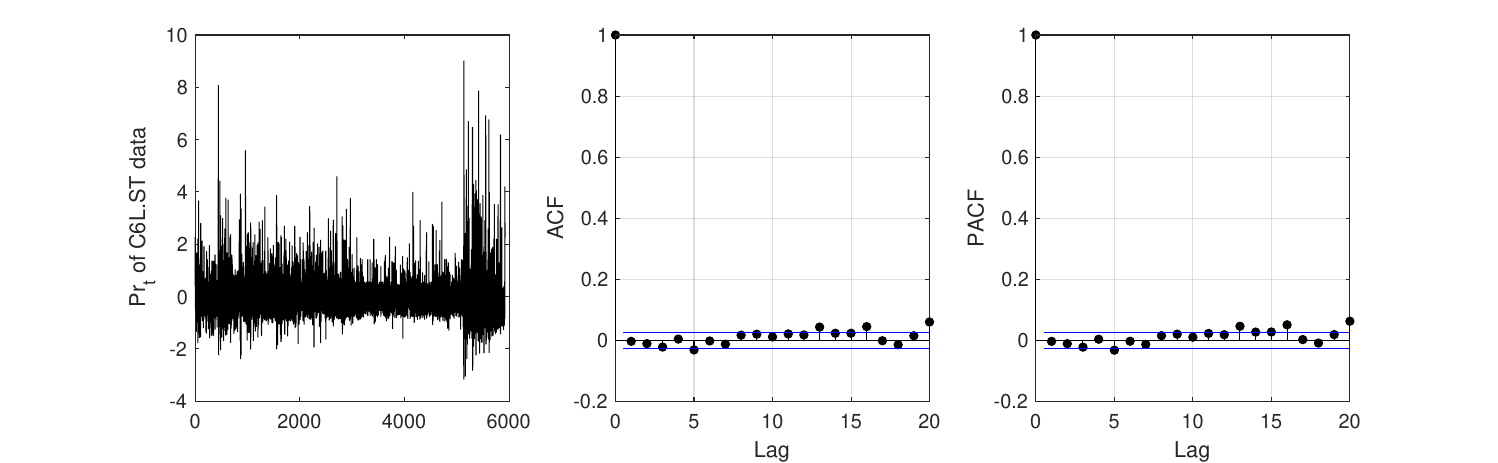}
\end{adjustwidth}
\vspace{-3mm}
\caption{Diagnostic checking plots of the fitted MCP-GCINAR model:\\ the traces, ACF and PACF plots of Pearson residual.}
\label{c6l_pr}
\end{figure}

From Table \ref{real_data}, it can be seen that the analysis results also show that the stock volumes of four airline groups have been relatively stable in the past 20 years or so, because a maximum of six change-points are detected in the large amount of data in each group.
As for the estimated model, the SE of estimated parameters in most estimated models is significant, except ``0753.HK" estimated model.
For this data set, some other models, such as integer-valued autoregressive conditional heteroskedastic (INGARCH) models, should be considered to fit this data in future studies.
Furthermore, it is discernible that the RMS outcome for 0753.HK is large, and its Pr$_{\rm Var}$ value  is relatively large.
This could be attributed to the presence of additional types of outliers in the segmented data of 0753.HK, such as additive outliers, thereby affecting the analytical results.
The diagnostic checking plots in Figure \ref{c6l_pr} display the Pearson residuals of Model MCP-GCINAR for C6L.SI dataset. These plots indicate that most lagged values of the ACF and PACF fall within the boundaries of the blue lines, suggesting a white noise characteristic for the residuals.
To further support this claim, we conduct the Ljung-Box (LB) test on the series of Pearson residuals. This test is performed with delay orders ranging from 1 to 10. Interestingly, all corresponding $p$-values are found to be greater than 0.05, suggesting that the series of Pearson residuals exhibits characteristics of white noise.

Next, the following international event that may have an impact on the airline groups is analyzed for the common change-points obtained.\\
$\bullet$
$\hat{\tau}=5103$ in C6L.SI data (February 10, 2020);
$\hat{\tau}=3701$ in 0753.HK data (December 6, 2019);
$\hat{\tau}=5886$ in LHA data (November 21, 2019);
$\hat{\tau}=3480$ in UAL data (December 20, 2019).
Since the outbreak of COVID-19 at the end of 2019, it has brought a great impact on various industries around the world, especially on the airline industry, which can be said to be unprecedented, which is also clearly reflected in the stock trading volume data.
From the data analysis, it can be seen that before the outbreak of COVID-19, the stock trading volume of airlines had remained at a relatively stable level.
However, since the beginning of COVID-19, the airline stock trading volume has almost all seen the most significant change-point since going public. To figure out why,
first of all, the epidemic led many governments to implement strict travel restrictions. Most flights were canceled and passenger demand dropped significantly, which led to the decline of airline stock prices.
A lack of confidence in the airline's future has led many investors to sell these shares, which is the main reason for the sharp increase in trading volume.
Second, airlines are under considerable financial pressure due to the rising costs of prolonged grounding, epidemic prevention and control and flight resumption, which also affect the performance of their stocks.
Finally, COVID-19's impact on the global economy has had a negative impact. As a result, airlines have been severely hit and may face the risk of bankruptcy, affecting investor confidence in the future of airlines and stock trading behavior.
\begin{table}[H]
\scriptsize
	\centering
	\caption{Summary of the change-points estimate results implemented by the LRSM  based on four airline groups data set. }\label{real_data}
	\begin{tabular}{*{6}{c}r*{3}{c}}
\toprule
&\multicolumn{7}{c}{LRSM}&\multicolumn{2}{c}{CIs}\\\cmidrule(lr){2-8}\cmidrule(lr){9-10}
Group&$\hat{m}$&$\hat{\tau}$&Date&MDL&RMS&Pr$_{\rm m}$&Pr$_{\rm Var}$&CI-PBA&CI-BBA($n_b$)\\\cmidrule(lr){1-10}
C6L.SI&1&5103&2020/2/10&$-$4438.7116&2.2851 &0.0015 &0.8932 &[5082,5118]&[5074,5115]($n_b=$60)\vspace{1mm}\\
0753.HK&6&420&2006/8/25&$-$180550.5027&12.2339 &0.0007 &1.6302 &[416,425]&[407,451.5]($n_b=$108)\\
&&994&2008/12/22&&&&&[989,1000]&[971,1036]($n_b=$108)\\
&&1934&2012/10/8&&&&&[1914,1954]&[1843,2158]($n_b=$432)\\
&&3175&2017/10/19&&&&&[3160,3198]&[3167,3315]($n_b=$288)\\
&&3701&2019/12/6&&&&&[3704,3757]&[3697,3755]($n_b=$328)\\
&&4030&2021/4/12&&&&&[3985,4027]&[3879,4015]($n_b=$168)\vspace{1mm}\\
LHA.DE&3&1159&2001/5/25&$-$18682.851&6.1740 &-0.0004 &1.2861 &[1015,1145]&[1008,1149]($n_b=$204)\\
&&5886&2019/11/21&&&&&[5799,5933]&[5701,6066]($n_b=$186)\\
&&6495&2022/4/20&&&&&[6500,6560]&[6501,6555]($n_b=$60)\vspace{1mm}\\
UAL&1&3480&2019/12/20&$-$54833.6158&7.3444 &-0.0002 &1.5487 &[3370,3552]&[3486,3660]($n_b=$362)\\
\bottomrule
	\end{tabular}
\end{table}
\begin{landscape}
~\\
The corresponding estimated model based on `C6L.SI' data set is as follows:
\begin{align}\label{estimate_C6L}
{\rm E}(X_{t}|\mathcal{F}_{t-1})=
\left\{
\begin{array}{lr}
0.3343X_{t-1,1}+0.1566X_{t-2,1}+0.0575X_{t-3,1}+0.0486X_{t-4,1}+0.0419X_{t-5,1}+0.0482X_{t-6,1}+0.6690&0<t\leq 5103,\\
(0.0192)~~~~~~~~~(0.0217)~~~~~~~~(0.0175)~~~~~~~~~~(0.0169)~~~~~~~~~(0.0168)~~~~~~~~~~(0.0157)~~~~~~~~~(0.0395)&\\
0.4335X_{t-5104,2}+0.0861X_{t-5105,2}+0.0266X_{t-5106,2}+0.1095X_{t-5107,2}+0.0988X_{t-5108,2}+1.8154&5103<t\leq 5927.\\
(0.0553)~~~~~~~~~(0.0426)~~~~~~~~(0.0385)~~~~~~~~~~(0.0396)~~~~~~~~~(0.0399)~~~~~~~~~~(0.2669)&\\
\end{array}\right.
\end{align}
where in parentheses are the standard errors (SE) of the estimators obtained from the square roots of the elements in the diagonal of the matrix $\hat{\bm{\Sigma_j}}=\bm{\hat{J}_j^{-1}}\bm{\hat{I}_j}\bm{\hat{J}_j^{-1}}$ computed on each segment $j$, with $\bm{\hat{J}_j}$ and $\bm{\hat{I}_j}$ are given by (\ref{SE1}) - (\ref{SE2}).
\begin{figure}[H]
\begin{center}
 \includegraphics[width=10.5in,height=3in]{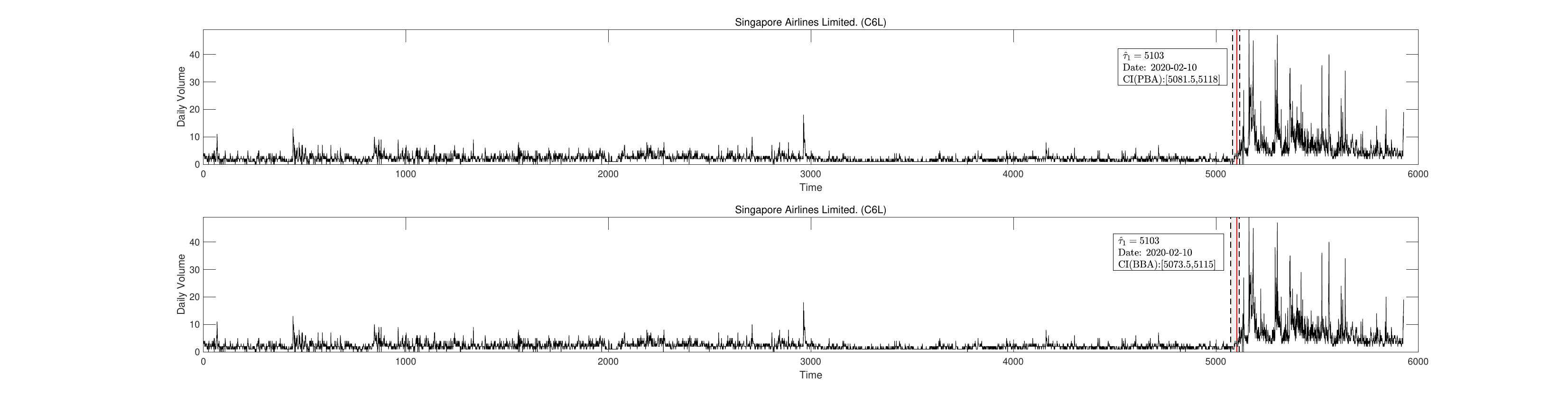}
\end{center}
\vspace{-6mm}
\caption{Sample path of the C6L.SI daily trading volume and estimate results given by the LRSM.\protect\\ The red line is the LRSM change-points estimate and the shaded regions are the corresponding PBA and BBA CIs.}\label{estimate_C6L_f}
\end{figure}
\end{landscape}

\section{Conclusion and discussion}\label{Set7}
In this paper, we propose a three-step LRSM, which provides a computationally valid and theoretically justified methods for change-points inference in the MCP-GCINAR process.
We infer that the computational complexity of the LRSM is $O((\log n)^3 n)$.
Simulation results and real data analysis results show that,
the LRSM with usual window parameter $h$ performs well in samples of long-time series with few and even change-points, and is robust when the assumption of window $h$ is violated.
In contrast, the LRSM with the multiple window parameter $h_{mix}$ performs well in short-time series with large and dense change-points, at the cost of high computational cost.
Furthermore, we demonstrated the asymptotic distribution of the change-points estimates.
The approximation distribution and two bootstrap procedures, parametric bootstrap
and block bootstrap, to approximate the finite sample distribution of change-point
estimates, and hence construct the CIs for the change-points.
Nevertheless, with proper modifications to the assumptions of the LRSM, it is possible to extend it to more generalized integer-value time series models, such as INGARCH models, integer-value moving average (INMA) models.
We will leave the above issues as our future work.
\section*{Acknowledgements}
This work is supported by National Natural Science Foundation of China (No.12271231).
\section*{Appendix}
We first state the following lemmas and their proofs, which are required to prove theorems in the LRSM.
Note that $\bm{\theta}$ corresponds one-to-one to the order $p$, to lighten notation, denote $\ell_t(\bm{\theta})=\ell_t(\bm{\theta},p)$ and $\xi_{t}(\bm{\theta})=\xi_{t}(\bm{\theta},p|{\bm X}_{t-1})$.
\begin{lemma}\label{lemma1}
For the $j$th change-point $\tau_j^0$, the scan statistic $S_h(\tau_j^0)$
\begin{align*}
S_h(\tau_j^0)=&\frac{1}{h}L_{h}(\tau_j^0-h,\hat{\bm{\theta}}_j,p_j)+\frac{1}{h}L_{h}(\tau_j^0,\hat{\bm{\theta}}_{j+1},p_{j+1})-\frac{1}{h}L_{2h} (\tau_j^0-h,\hat{\bm{\theta}}_{j,j+1},p_{j,j+1})\\
=&\frac{1}{h}\sum\limits_{t=\tau_j^0-h+1}^{\tau_j^0}\ell_{t,\bm{\theta}_{j}^0}(\hat{\bm{\theta}}_{j}) +\frac{1}{h}\sum\limits_{t=\tau_j^0+1}^{\tau_j^0+h}\ell_{t,\bm{\theta}_{j+1}^0}(\hat{\bm{\theta}}_{j+1}) -\frac{1}{h}\Big[\sum\limits_{t=\tau_j^0-h+1}^{\tau_j^0}\ell_{t,\bm{\theta}_{j}^0}(\hat{\bm{\theta}}_{j,j+1}) +\sum\limits_{t=\tau_j^0+1}^{\tau_j^0+h}\ell_{t,\bm{\theta}_{j+1}^0}(\hat{\bm{\theta}}_{j,j+1})\Big]\\
\xrightarrow{p}&\sum\limits_{i=0}^1{\rm E}\Big\{\ell_{t,\bm{\theta}_{j+i}^0}(\bm{\theta}_{j+i}^0)-\ell_{t,\bm{\theta}_{j+i}^0}(\bm{\theta}_{j,j+1})\Big\}\\
=&\sum\limits_{i=0}^1{\rm E}\Big\{X_t\log\frac{\xi_{t}(\bm{\theta}_{j+i}^0)}{\xi_{t}(\bm{\theta}_{j,j+1})}-
\big[\xi_{t}(\bm{\theta}_{j+i}^0)-\xi_{t}(\bm{\theta}_{j,j+1})\big]\Big\}\triangleq g_j>0.
\end{align*}
where
\begin{align*}
\hat{\bm{\theta}}_{j,j+1}&=\arg\max\limits_{\bm{\theta}\in\bm{\Theta}}\frac{1}{h} \Big[\sum\limits_{t=\tau_j^0-h+1}^{\tau_j^0}\ell_{t,\bm{\theta}_j^0}(\bm{\theta})+ \sum\limits_{t=\tau_j^0+1}^{\tau_j^0+h}\ell_{t,\bm{\theta}_{j+1}^0}(\bm{\theta})\Big],\\
\bm{\theta}_{j,j+1}&=\arg\max\limits_{\bm{\theta}\in\bm{\Theta}}\Big[{\rm E}\big(\ell_{t,\bm{\theta}_j^0}(\bm{\theta})\big)+{\rm E}\big(\ell_{t,\bm{\theta}_{j+1}^0}(\bm{\theta})\big)\Big].
\end{align*}
and $\ell_{t,\bm{\theta}_{s}^0}(\bm{\theta})$ represents $\ell_{t}(\bm{\theta},p)$ with the true parameter $\bm{\theta}_{s}^0$ .
\end{lemma}
$\mathbf{Proof~of~Lemma~\ref{lemma1}.}$ We first have
\begin{align*}
\hat{\bm{\theta}}_{j,j+1}&=\arg\max\limits_{\bm{\theta}\in\bm{\Theta}}\frac{1}{h} \Big[\sum\limits_{t=\tau_j^0-h+1}^{\tau_j^0}\ell_{t,\bm{\theta}_j^0}(\bm{\theta})+ \sum\limits_{t=\tau_j^0+1}^{\tau_j^0+h}\ell_{t,\bm{\theta}_{j+1}^0}(\bm{\theta})\Big],\\
& =\arg\max\limits_{\bm{\theta}\in\bm{\Theta}}\Big[{\rm E}\big(\ell_{t,\bm{\theta}_j^0}(\bm{\theta})\big)+ {\rm E}\big(\ell_{t,\bm{\theta}_{j+1}^0}(\bm{\theta})\big)+o_p(1)\Big]\xrightarrow{p}\bm{\theta}_{j,j+1},~~\text{as}~~h\rightarrow \infty.
\end{align*}
Combine $\hat{\bm{\theta}}_{j}\xrightarrow{p} \bm{\theta}_j^0$ and $\hat{\bm{\theta}}_{j+1}\xrightarrow{p} \bm{\theta}_{j+1}^0$, after some simple transformations
\begin{align}
S_h(\tau_j^0)&=\frac{1}{h}L_{h}(\tau_j^0-h,\hat{\bm{\theta}}_j,p_j)+\frac{1}{h}L_{h}(\tau_j^0,\hat{\bm{\theta}}_{j+1},p_{j+1})-\frac{1}{h}L_{2h} (\tau_j^0-h,\hat{\bm{\theta}}_{j,j+1},p_{j,j+1})\nonumber\\
&\xrightarrow{p}\sum\limits_{i=0}^1{\rm E}\Big\{\ell_{t,\bm{\theta}_{j+i}^0}(\bm{\theta}_{j+i}^0)-\ell_{t,\bm{\theta}_{j+i}^0}(\bm{\theta}_{j,j+1})\Big\}\label{lemma1.1}\\
&=\sum\limits_{i=0}^1{\rm E}\Big\{X_t\log\frac{\xi_{t}(\bm{\theta}_{j+i}^0)}{\xi_{t}(\bm{\theta}_{j,j+1})}-
\big[\xi_{t}(\bm{\theta}_{j+i}^0)-\xi_{t}(\bm{\theta}_{j,j+1})\big]\Big\}= g_j.\nonumber
\end{align}
Next, to show $g_j>0$. By the definition of maximum likelihood estimate, for the first part ($i=0$) in eq.(\ref{lemma1.1}),
\begin{align*}
{\rm E}\big[\ell_{t,\bm{\theta}_{j}^0}(\bm{\theta}_{j}^0)-\ell_{t,\bm{\theta}_{j}^0}(\bm{\theta}_{j,j+1})\big]\geq0,
\end{align*}
the equality sign is true if and only if $\bm{\theta}_{j,j+1}=\bm{\theta}_{j}^0$. Similarly, for the second part ($i=1$),
\begin{align*}
{\rm E}\big[\ell_{t,\bm{\theta}_{j+1}^0}(\bm{\theta}_{j+1}^0)-\ell_{t,\bm{\theta}_{j+1}^0}(\bm{\theta}_{j,j+1})\big]\geq0,
\end{align*}
the equality sign is true if and only if $\bm{\theta}_{j,j+1}=\bm{\theta}_{j+1}^0$. Note that $\bm{\theta}_{j}^0\neq\bm{\theta}_{j+1}^0$, which  means that both parts cannot be equal at the same time, that is $g_j>0$, and the proof is completed.
~\\

\begin{lemma}\label{lemma2}
Let $\{X_t\}$ be a piecewise stationary MCP-GCINAR process defined in (\ref{CGINAR}),
and $Y_t(\bm{\theta})=\ell_t(\bm{\theta})-{\rm E}\big(\ell_t(\bm{\theta})\big)$. 
For all $\bm{\theta}\in\bm{\Theta}$, if ${\rm E}(X_{t})^{2k+2\epsilon_X}<\infty$ for some $\epsilon_X>0$,
then $\sup_{\bm{\theta}\in\bm{\Theta}}{\rm E}|\ell_t(\bm{\theta})|^{k+\epsilon_X}<\infty$.
\end{lemma}
$\mathbf{Proof~of~Lemma~\ref{lemma2}.}$ By the definition of $\ell_t(\bm{\theta_j})$, we have
\begin{align*}
{\rm E}|\ell_t(\bm{\theta_j})|^{k+\epsilon_X}&={\rm E}|X_{t,j}\log \xi_{t}(\bm{\theta}_j)-\xi_{t}(\bm{\theta}_j)|^{k+\epsilon_X}\\
&\leq{\rm E}|X_{t}\log \xi_{t}(\bm{\theta_j})+\xi_{t}(\bm{\theta_j})|^{k+\epsilon_X}.
\end{align*}
Clearly, there exist a suitable constant $0<C_1<\infty$ satisfy
\begin{align*}
{\rm E}|\ell_t(\bm{\theta_j},p_j)|^{k+\epsilon_X}\leq C_1{\rm E}|X_{t}^2|^{k+\epsilon_X}.
\end{align*}
That is, if ${\rm E}(X_{t})^{2k+2\epsilon_X}<\infty$, there is
\begin{align*}
\sup_{\bm{\theta}\in\bm{\Theta}}{\rm E}|\ell_t(\bm{\theta})|^{k+\epsilon_X}<\infty~\text{and}~\sup_{\bm{\theta}\in\bm{\Theta}}{\rm E}|Y_t(\bm{\theta})|^{k+\epsilon_X}<\infty.
\end{align*}
The proof of Lemma \ref{lemma2} is completed.
~\\

\begin{lemma}\label{lemma3}
For any $\epsilon>0$, there exists a positive integer $C_2$ such that for any $h>C_2$,
\begin{align*}
{\rm P}\Big(|S_h(t)|>\epsilon\Big)\leq6\exp(-\frac{1}{4}h^{1/3}\epsilon^{2/3}),
\end{align*}
for all $t$ such that $W_t(h)$ does not contain any change-point.
\end{lemma}
$\mathbf{Proof~of~Lemma~\ref{lemma3}.}$
Since the scanning window $W_t(h)$ has no change point, we can assume that all data in the window comes from the segment specified by $\bm{\theta}^0$.
Hence, $S_h(t)$ can be written as
\begin{align}
S_h(t)&=\frac{1}{h}\sum\limits_{s=t-h+1}^{t}\big[\ell_s(\hat{\bm{\theta}}_1)-\ell_s(\bm{\theta}^0)\big] +\frac{1}{h}\sum\limits_{s=t+1}^{t+h}\big[\ell_s(\hat{\bm{\theta}}_2)-\ell_s(\bm{\theta}^0)\big] -\frac{1}{h}\sum\limits_{s=t-h+1}^{t+h}\big[\ell_s(\hat{\bm{\theta}})-\ell_s(\bm{\theta}^0)\big]\nonumber\\
&={\rm I}+{\rm II}+{\rm III},\label{lemma3.1}
\end{align}
where $\hat{\bm{\theta}}_1$, $\hat{\bm{\theta}}_2$ and $\hat{\bm{\theta}}$ and the PQML estimates of the parameter $\bm{\theta}$ in the left half, right half, and the entire scanning window, respectively.
For the third part (${\rm III}$) in eq.(\ref{lemma3.1}),
\begin{align*}
\frac{1}{h}\sum\limits_{s=t-h+1}^{t+h}\big[\ell_s(\hat{\bm{\theta}})-\ell_s(\bm{\theta}^0)\big]
&=\frac{1}{h}\sum\limits_{s=t-h+1}^{t+h}\big[\ell_s(\hat{\bm{\theta}})-{\rm E}\big(\ell_s(\hat{\bm{\theta}})\big)\big]
-\frac{1}{h}\sum\limits_{s=t-h+1}^{t+h}\big[\ell_s(\bm{\theta}^0)-{\rm E}\big(\ell_s(\bm{\theta}^0)\big)\big]\nonumber\\
&+2\big[{\rm E}\big(\ell_s(\hat{\bm{\theta}})\big)-{\rm E}\big(\ell_s(\bm{\theta}^0)\big)\big].
\end{align*}
To simplify notation, let $Y_s(\bm{\theta},p)=\ell_s(\bm{\theta})-{\rm E}\big(\ell_s(\bm{\theta})\big)$.
Within any segment, $\ell_s(\bm{\theta})$ is a measurable and Lipschitz continuous function with respect to $\{X_t\}$.
under the assumption H.\ref{H7}, there exists a $C_2^{(1)}$ such that ${\rm E}\big(e^{|Y_s(\bm{\theta})|}\big)\leq C_2^{(1)}$,
and $Y_s(\bm{\theta})$ is obviously a finite sequence of martingale differences for $s=t-h+1,...,t+h$ and $\bm{\theta}\in\bm{\Theta}$.
Then, following the Theorem 3.2 of \cite{Lesigne2001}, for any $\epsilon>0$, there exists a positive integer $C_2^{(2)}$ depending only on $C_2^{(1)}$ and $\epsilon$, such that,
for any $h>C_2^{(2)}$ and $\bm{\theta}\in\bm{\Theta}$,
\begin{align}\label{lemma3.2}
{\rm P}\Big(\Big|\frac{1}{h}\sum\limits_{s=t-h+1}^{t+h}\big[\ell_s(\bm{\theta})-{\rm E}\big(\ell_s(\bm{\theta})\big)\big]\Big|>\frac{\epsilon}{3}\Big)\leq \exp(-\frac{1}{4}h^{1/3}\epsilon^{2/3}).
\end{align}
Moreover, $\ell_s(\bm{\theta})$ is the uniform integrability for all $\bm{\theta}\in\bm{\Theta}$, combine $\hat{\bm{\theta}}\xrightarrow{p} \bm{\theta}^0$, we have that, for any $\epsilon>0$, there exists a constant $C_2^{(3)}>0$ such that, for any $h>C_2^{(3)}$,
\begin{align}\label{lemma3.3}
2\Big|{\rm E}\big(\ell_s(\hat{\bm{\theta}})\big)-{\rm E}\big(\ell_s(\bm{\theta}^0)\big)\Big|<\frac{\epsilon}{3}
\end{align}
Thus, applying (\ref{lemma3.2}) to $\hat{\bm{\theta}}$ and $\bm{\theta}^0$, there exists a constant $C_2^{(4)}>0$ such that, for any $h>C_2^{(4)}$,
\begin{align*}
{\rm P}\Big(\Big|\frac{1}{h}\sum\limits_{s=t-h+1}^{t+h}\big[\ell_s(\hat{\bm{\theta}})-{\rm E}\big(\ell_s(\hat{\bm{\theta}})\big)\big]\Big|>\frac{\epsilon}{3}\Big)\leq \exp(-\frac{1}{4}h^{1/3}\epsilon^{2/3}),\\
{\rm P}\Big(\Big|\frac{1}{h}\sum\limits_{s=t-h+1}^{t+h}\big[\ell_s(\bm{\theta}^0)-{\rm E}\big(\ell_s(\bm{\theta}^0)\big)\big]\Big|>\frac{\epsilon}{3}\Big)\leq \exp(-\frac{1}{4}h^{1/3}\epsilon^{2/3}).
\end{align*}
Then, together with (\ref{lemma3.3}),
we obtain that, for any $\epsilon>0$, there exists a constant $C_2^{(5)}=\max\{C_2^{(3)},C_2^{(4)}\}$ such that, for any $h>C_2^{(5)}$,
\begin{align}\label{lemma3.4}
{\rm P}\Big(\Big|\frac{1}{h}\sum\limits_{s=t-h+1}^{t+h}\big[\ell_s(\hat{\bm{\theta}})-\ell_s(\bm{\theta}^0)\big]\Big|>\epsilon\Big)
\leq 2\exp(-\frac{1}{4}h^{1/3}\epsilon^{2/3}).
\end{align}
Similarly exponential inequalities hold for the other two parts (${\rm I}$) and (${\rm II}$) in eq.(\ref{lemma3.1}), and the proof of Lemma \ref{lemma3} is completed.
~\\

\begin{lemma}\label{lemma4}
For any $\epsilon>0$, there exists a positive integer $C_3$ such that for any $h>C_3$,
\begin{align*}
{\rm P}\Big(|S_h(\tau_j^0)-g_j|>\epsilon\Big)\leq 22\exp(-\frac{1}{4}h^{1/3}\epsilon^{2/3}),
\end{align*}
for all $j=1,...,m_0$.
\end{lemma}
$\mathbf{Proof~of~Lemma~\ref{lemma4}.}$ Using the notations in Lemma \ref{lemma1}, there is
\begin{align*}
&S_h(\tau_j^0)-g_j\\
=&\Big[\frac{1}{h}\sum\limits_{t=\tau_j^0-h+1}^{\tau_j^0}\ell_{t,\bm{\theta}_{j}^0}(\hat{\bm{\theta}}_{j})
-{\rm E}\ell_{t,\bm{\theta}_{j}^0}(\bm{\theta}_{j}^0)\Big] +\Big[\frac{1}{h}\sum\limits_{t=\tau_j^0+1}^{\tau_j^0+h}\ell_{t,\bm{\theta}_{j+1}^0}(\hat{\bm{\theta}}_{j+1})
-{\rm E}\ell_{t,\bm{\theta}_{j+1}^0}(\bm{\theta}_{j+1}^0)\Big]\\ &-\Big[\frac{1}{h}\sum\limits_{t=\tau_j^0-h+1}^{\tau_j^0}\Big(\ell_{t,\bm{\theta}_{j}^0}(\hat{\bm{\theta}}_{j,j+1}) +\ell_{t+h,\bm{\theta}_{j+1}^0}(\hat{\bm{\theta}}_{j,j+1})\Big)
-{\rm E}\Big(\ell_{t,\bm{\theta}_{j}^0}(\bm{\theta}_{j,j+1})
+\ell_{t,\bm{\theta}_{j+1}^0}(\bm{\theta}_{j,j+1})\Big)\Big]\\
=&\Big[\frac{1}{h}\sum\limits_{t=\tau_j^0-h+1}^{\tau_j^0}\big(\ell_{t,\bm{\theta}_{j}^0}(\hat{\bm{\theta}}_{j})
-\ell_{t,\bm{\theta}_{j}^0}(\bm{\theta}_{j}^0)\big)\Big]
+\Big[\frac{1}{h}\sum\limits_{t=\tau_j^0-h+1}^{\tau_j^0}\ell_{t,\bm{\theta}_{j}^0}(\bm{\theta}_{j}^0)
-{\rm E}\ell_{t,\bm{\theta}_{j}^0}(\bm{\theta}_{j}^0)\Big]\\
&+\Big[\frac{1}{h}\sum\limits_{t=\tau_j^0+1}^{\tau_j^0+h}\big(\ell_{t,\bm{\theta}_{j+1}^0}(\hat{\bm{\theta}}_{j+1})
-\ell_{t,\bm{\theta}_{j+1}^0}(\bm{\theta}_{j+1}^0)\big)\Big]
+\Big[\frac{1}{h}\sum\limits_{t=\tau_j^0+1}^{\tau_j^0+h}\ell_{t,\bm{\theta}_{j+1}^0}(\bm{\theta}_{j+1}^0)
-{\rm E}\ell_{t,\bm{\theta}_{j+1}^0}(\bm{\theta}_{j+1}^0)\Big]\\
&+\frac{1}{h}\sum\limits_{t=\tau_j^0-h+1}^{\tau_j^0}\Big(\big(\ell_{t,\bm{\theta}_{j}^0}(\hat{\bm{\theta}}_{j,j+1}) +\ell_{t+h,\bm{\theta}_{j+1}^0}(\hat{\bm{\theta}}_{j,j+1})\big)
-\big(\ell_{t,\bm{\theta}_{j}^0}(\bm{\theta}_{j,j+1}) +\ell_{t+h,\bm{\theta}_{j+1}^0}(\bm{\theta}_{j,j+1})\big)\Big)\\
&+\Big[\frac{1}{h}\sum\limits_{t=\tau_j^0-h+1}^{\tau_j^0}\ell_{t,\bm{\theta}_{j}^0}(\bm{\theta}_{j,j+1})
-{\rm E}\ell_{t,\bm{\theta}_{j}^0}(\bm{\theta}_{j,j+1})\Big]
+\Big[\frac{1}{h}\sum\limits_{t=\tau_j^0+1}^{\tau_j^0+h}\ell_{t,\bm{\theta}_{j+1}^0}(\bm{\theta}_{j,j+1})
-{\rm E}\ell_{t,\bm{\theta}_{j+1}^0}(\bm{\theta}_{j,j+1})\Big]\\
=&{\rm I} + {\rm II }+{\rm III} + {\rm IV }
\end{align*}
For the parts ${\rm I}$ and ${\rm II }$, according to equations (\ref{lemma3.2}) and (\ref{lemma3.4}) in Lemma \ref{lemma3},
we have that for any $\epsilon>0$, there exists a constant $C_3^{(1)}$ such that, for any $h>C_3^{(1)}$,
\begin{align*}
&{\rm P}\Big(\Big|\frac{1}{h}\sum\limits_{t=\tau_j^0-h+1}^{\tau_j^0}\big[\ell_{t,\bm{\theta}_{j}^0}(\hat{\bm{\theta}}_{j})
-\ell_{t,\bm{\theta}_{j}^0}(\bm{\theta}_{j}^0)\big]\Big|>\epsilon\Big)
+{\rm P}\Big(\Big|\frac{1}{h}\sum\limits_{t=\tau_j^0-h+1}^{\tau_j^0}\ell_{t,\bm{\theta}_{j}^0}(\bm{\theta}_{j}^0)
-{\rm E}\ell_{t,\bm{\theta}_{j}^0}(\bm{\theta}_{j}^0)\Big|>\epsilon\Big)\\
&\leq 5\exp(-\frac{1}{4}h^{1/3}\epsilon^{2/3}).
\end{align*}
and
\begin{align}
&{\rm P}\Big(\Big|\frac{1}{h}\sum\limits_{t=\tau_j^0+1}^{\tau_j^0+h}\big[\ell_{t,\bm{\theta}_{j+1}^0}(\hat{\bm{\theta}}_{j+1})
-\ell_{t,\bm{\theta}_{j+1}^0}(\bm{\theta}_{j+1}^0)\big]\Big|>\epsilon\Big)
+{\rm P}\Big(\Big|\frac{1}{h}\sum\limits_{t=\tau_j^0+1}^{\tau_j^0+h}\ell_{t,\bm{\theta}_{j+1}^0}(\bm{\theta}_{j+1}^0)
-{\rm E}\ell_{t,\bm{\theta}_{j+1}^0}(\bm{\theta}_{j+1}^0)\Big|>\epsilon\Big)\nonumber\\
&\leq 5\exp(-\frac{1}{4}h^{1/3}\epsilon^{2/3}).\label{lemma4.1}
\end{align}
By the definition of $\bm{\theta}_{j,j+1}$, we have
$$\frac{\partial}{\partial\bm{\theta}}{\rm E}\big(\ell_{t,\bm{\theta}_{j}^0}(\bm{\theta}) +\ell_{t+h,\bm{\theta}_{j+1}^0}(\bm{\theta})\big)\Big|_{\bm{\theta}=\bm{\theta}_{j,j+1}}=0.$$
Obviously, $\ell_{t,\bm{\theta}}(\bm{\theta})$ is twice continuously differentiable with respect to $\bm{\theta}$ almost surely, thus
$${\rm E}\frac{\partial}{\partial\bm{\theta}}\big(\ell_{t,\bm{\theta}_{j}^0}(\bm{\theta}_{j,j+1}) +\ell_{t+h,\bm{\theta}_{j+1}^0}(\bm{\theta}_{j,j+1})\big)=0.$$
That is, $\Big\{\frac{\partial}{\partial\bm{\theta}}\big(\ell_{t,\bm{\theta}_{j}^0}(\bm{\theta}_{j,j+1}) +\ell_{t+h,\bm{\theta}_{j+1}^0}(\bm{\theta}_{j,j+1})\big)\Big\}$ is a martingale difference sequence.
Similarly to the proof of Lemma \ref{lemma3}, we have that for any $\epsilon>0$, there exists a constant $C_3^{(2)}$ such that, for any $h>C_3^{(2)}$,
\begin{align}
&{\rm P}\Big(\Big|\frac{1}{h}\sum\limits_{t=\tau_j^0-h+1}^{\tau_j^0}\big[\big(\ell_{t,\bm{\theta}_{j}^0}(\hat{\bm{\theta}}_{j,j+1}) +\ell_{t+h,\bm{\theta}_{j+1}^0}(\hat{\bm{\theta}}_{j,j+1})\big)
-\big(\ell_{t,\bm{\theta}_{j}^0}(\bm{\theta}_{j,j+1}) +\ell_{t+h,\bm{\theta}_{j+1}^0}(\bm{\theta}_{j,j+1})\big)\big]\Big|>\epsilon\Big)\nonumber\\
&\leq 6\exp(-\frac{1}{4}h^{1/3}\epsilon^{2/3}).\label{lemma4.3}
\end{align}
Also, applying (\ref{lemma3.2}) to $\bm{\theta}_{j,j+1}$, for any $\epsilon>0$, there exists a constant $C_3^{(3)}$ such that, for any $h>C_3^{(3)}$,
\begin{align}\label{lemma4.4}
{\rm P}\Big(\Big|\frac{1}{h}\sum\limits_{t=\tau_j^0-h+1}^{\tau_j^0}\ell_{t,\bm{\theta}_{j}^0}(\bm{\theta}_{j,j+1})
-{\rm E}\ell_{t,\bm{\theta}_{j}^0}(\bm{\theta}_{j,j+1})\Big|>\epsilon\Big)\leq 3\exp(-\frac{1}{4}h^{1/3}\epsilon^{2/3}),
\end{align}
and
\begin{align}\label{lemma4.5}
{\rm P}\Big(\Big|\frac{1}{h}\sum\limits_{t=\tau_j^0+1}^{\tau_j^0+h}\ell_{t,\bm{\theta}_{j+1}^0}(\bm{\theta}_{j,j+1})
-{\rm E}\ell_{t,\bm{\theta}_{j+1}^0}(\bm{\theta}_{j,j+1})\Big|>\epsilon\Big)\leq 3\exp(-\frac{1}{4}h^{1/3}\epsilon^{2/3}).
\end{align}
Combining formulas (\ref{lemma4.1}) - (\ref{lemma4.5}), and the proof of Lemma \ref{lemma4} is completed.
~\\
~\\
The proof of Theorem~\ref{LRSM1_theory} is similar to the proof of Theorem 1 in \cite{YauZhao2016}.\\
$\mathbf{Proof~of~Theorem~\ref{LRSM1_theory}.}$
Let $A_t=$\{ some point in the $t$-th local-window is a local change-point estimate\} and $A=\bigcap_{t\in \mathcal{J}_0}A_t$.
If we can prove that ${\rm P}(A)\rightarrow1$ as $n\rightarrow\infty$, then the proof of Theorem \ref{LRSM1_theory} holds.
Let $\mathbb{Z}_n=\{1,2,...,n\}$. Define $\mathcal{E}=\mathbb{Z}_n\setminus\bigcap_{t\in \mathcal{J}_0}W_t(h)$ as the set of all points outside the $h$-neighbourhood of the true change-points.
A sufficient condition for the event $A$ to occur is that
\begin{align}\label{LRSM1_theory_1}
\min_{t\in \mathcal{J}_0}S_h(t)>\max_{t\in \mathcal{E}}S_h(t).
\end{align}
Let $g=\frac{1}{2}\min_{j=1,...,m_0}(g_j)$ where $g_j$s are defined in Lemma \ref{lemma1}.
Note that from (\ref{LRSM1_theory_1}), there is
\begin{align*}
{\rm P}(A) \geq{\rm P}\Big(\min_{t\in \mathcal{J}_0}S_h(t)>g>\max_{t\in \mathcal{E}}S_h(t)\Big).
\end{align*}
Hence, the proof of Theorem \ref{LRSM1_theory} is completed by proving the following two facts,\\
(i) ${\rm P}\Big(\min_{t\in \mathcal{J}_0}S_h(t)>g\Big)\rightarrow 1,$\\
(ii) ${\rm P}\Big(g>\max_{t\in \mathcal{E}}S_h(t)\Big)\rightarrow 1.$\\
For (i),
\begin{align*}
{\rm P}\Big(\min_{t\in \mathcal{J}_0}S_h(t)>g\Big)=1-{\rm P}\Big(\bigcup_{t\in \mathcal{J}_0}\{S_h(t)\leq g\}\Big)
\geq 1-\sum_{t\in \mathcal{J}_0}{\rm P}\big(S_h(t)\leq g\big),
\end{align*}
from Lemma \ref{lemma4} and the definition of $g$, it can be shown that
${\rm P}\big(S_h(t)\leq g\big)\leq 22\exp(-\frac{1}{4}h^{1/3}\epsilon^{2/3}),$ for all $t\in \mathcal{J}_0$.
Thus, set $h=d(\log n)^3$, for some $d>0$ and $m_0=O(1)$, we have
\begin{align*}
{\rm P}\Big(\min_{t\in \mathcal{J}_0}S_h(t)>g\Big)
\geq 1-22m_0\exp(-\frac{1}{4}h^{1/3}\epsilon^{2/3})\rightarrow 1.
\end{align*}
For (ii), note that when $t\in\mathcal{E}$, all observations in $S_h(t)$ belong to one stationary piece.
From Lemma \ref{lemma3}, ${\rm P}\Big(S_h(t)\geq g\Big)<6\exp(-\frac{1}{4}h^{1/3}\epsilon^{2/3})$ for all $t\in\mathcal{E}$.
Thus, set $h=d(\log n)^3$, for some $d>64/g^2$, we have
\begin{align*}
{\rm P}\Big(g>\max_{t\in \mathcal{E}}S_h(t)\Big)&=1-{\rm P}\Big(\bigcup_{t\in \mathcal{E}}\{S_h(t_j)\geq g\}\Big)
\geq 1-\sum_{j=1}^{m_0+1}(\tau_j^0-\tau_{j-1}^9)\big(S_h(t)\geq g\big),\\
&>1-6\exp(-\frac{1}{4}h^{1/3}\epsilon^{2/3})\rightarrow1
\end{align*}
where $t_j\in(\tau_{j-1}^0+h,\tau_{j}^0-h)$. To sum up, when $h=d(\log n)^3$ for some $d>64/g^2$,
there is ${\rm P}(A)\rightarrow1$, and the proof of Theorem \ref{LRSM1_theory} is completed.
~\\
~\\
$\mathbf{Proof~of~Theorem~\ref{LRSM2_theory}.}$
In \cite{Sheng2023}, it shows that, if $m_0$ is known, under the assumption ${\rm E}|X_{t,j}|^{2+\epsilon_X}<\infty$, estimates based on MDL principle are strongly consistent.
If $m_0$ is unknown, under the assumption ${\rm E}|X_{t,j}|^{4+\epsilon_X}<\infty$,
estimates are weakly consistent, further, under the assumption ${\rm E}|X_{t,j}|^{8+\epsilon_X}<\infty$, change-points estimates are strongly consistent.
~\\
~\\
$\mathbf{Proof~of~Theorem~\ref{LRSM3_theory}.}$
Let $G({\bm X}_t)=|X_{t}\log \xi_{t}(\bm{\theta_j})+\xi_{t}(\bm{\theta_j})|$. Clearly, $G({\bm X}_t)$ is an integrable function and
$|\ell_t(\bm{\theta}_j)|\leq G({\bm X}_t)$ for all $\bm{\theta}_j\in\bm{\Theta}_j$.
Furthermore, from Lemma \ref{lemma2}, ${\rm E}\big(G({\bm X}_t)\big)<\infty$.
By using the uniform law of large number in \cite{Jennrich1969}, we have as $h\rightarrow \infty$,
$\frac{1}{h}\sum_{t=1}^h \ell_t(\bm{\theta}_j)$ converges uniformly to ${\rm E}_{{\bm\theta}_j^0}\big(\ell_t(\bm{\theta}_j)\big)$,
for any $\bm{\theta}_j\in\bm{\Theta}_j$.
The remaining part of the proof is the same as that of Theorem 3 in \cite{YauZhao2016} and we omit it.
~\\
~\\
$\mathbf{Proof~of~Theorem~\ref{asym}.}$ The proof of Theorem~\ref{asym} is similar to the Theorem 3 of \cite{Cui2021}, and we omit it.
~\\
~\\
\begin{lemma}\label{lemma5}
Let
\begin{align}
g_{t}(\bm{\theta}_1,\bm{\theta}_2,\bm{X}_t)=&{\rm sgn}(t)\big(\ell_t(\bm{\theta}_{1},\bm{X}_t) -\ell_t(\bm{\theta}_{2},\bm{X}_t) \big),\label{lemma5.1}\\
=&{\rm sgn}(t)\Bigg\{\Big[X_{t}\log\frac{\sum\limits_{i=1}^{p}\beta_{i,1} X_{t-i}+\beta_{0,1}}{\sum\limits_{i=1}^{p}\beta_{i,2} X_{t-i}+\beta_{0,2}}-\big(\sum\limits_{i=1}^{p}(\beta_{i,1}-\beta_{i,2}) X_{t-i}+\beta_{0,1}-\beta_{0,2}\big)\Big]\Bigg\}\nonumber
\end{align}
where $\bm{\theta}_1=(\beta_{0,1},...,\beta_{p,1})$ and $\bm{\theta}_2=(\beta_{0,2},...,\beta_{p,2})$ are the interior points of the compact space
$\bm{\Theta}(p)=[\delta,\tilde{\delta}]\times[0,1-\delta]^{p}\cap\mathbb{M}$, where $\mathbb{M}=\{0\leq\sum_{k=1}^{p}\beta_{k}\leq1-\delta<1\}$, $\delta$ and $\tilde{\delta}$ are finite positive constants with $\delta$ approaching $0$ and $\tilde{\delta}<+\infty$,
and ${\rm sgn}(t)=1$ with $t>0$, ${\rm sgn}(t)=0$ with $t=0$, ${\rm sgn}(t)=-1$ with $t<0$.
Then the function $g_{t}(\bm{\theta}_1,\bm{\theta}_2,\bm{X}_t)$ about $\bm{X}_t$ has partial derivatives, which satisfy a Lipschitz condition, that is, the derivative
$$\frac{\partial^{p+1} g_{t}(\bm{\theta}_1,\bm{\theta}_2,\bm{X}_t)}{\partial X_t\partial X_{t-1}...\partial X_{t-p}}$$
is Lipschitz.
\end{lemma}

$\mathbf{Proof~of~Lemma~\ref{lemma5}.}$
For any $\bm{\theta}_1, \bm{\theta}_2\in \bm{\Theta}(p)$, $\bm{X}_t\in [0,+\infty)^{p+1}$, there exist a constant $0<C_5^{(1)}<\infty$, such that
\begin{align*}
\Bigg|\frac{\partial^{p+1} g_{t}(\bm{\theta}_1,\bm{\theta}_2,\bm{X}_t)}{\partial X_t\partial X_{t-1}...\partial X_{t-p}}\Bigg| &=\Bigg|(-1)^p\Big(\frac{\prod_{i=1}^p\beta_{i,1}}{\sum_{i=1}^{p}\beta_{i,1} X_{t-i} +\beta_{0,1}}
-\frac{\prod_{i=1}^p\beta_{i,2}}{\sum_{i=1}^{p}\beta_{i,2} X_{t-i}+\beta_{0,2}}\Big)\Bigg|\\
&< \beta_{0,1}^{-1}+\beta_{0,2}^{-1}<C_5^{(1)},
\end{align*}
and the Lemma \ref{lemma5} is clearly true.
~\\
~\\
$\mathbf{Proof~of~Theorem~\ref{LRSM5_theory}.}$
Without loss of generality, we consider the segment before and after the $j$th change-point, and the other change-points are similarly.
Denote $\mathcal{L}(\bm{X})$, $\widetilde{\mathcal{L}}(\bm{X})$, and $\mathcal{L}^{*}(\bm{X})$
as the distribution of random variable $\bm{X}$ under probability measures ${\rm P}$, $\widetilde{{\rm P}}$, and ${\rm P}^{*}$, respectively.
We first prove the assertion (\ref{theorem5.1}) in Theorem \ref{LRSM5_theory}, that is, conditional on the sample $\{X_t\}_{t=\hat{\tau}_{j-1}^{(3)}+1}^{\hat{\tau}_{j+1}^{(3)}}$, for any finite $n_p\in \mathbb{N}$, there is
\begin{align}
\widetilde{\mathcal{L}}(\bm{W}_j)&\triangleq \widetilde{\mathcal{L}}(\widetilde{W}_{j,-n_p},\widetilde{W}_{j,-n_p+1},...,\widetilde{W}_{j,n_p})\nonumber\\
&\xrightarrow{d} \mathcal{L}(W_{j,-n_p},W_{j,-n_p+1},...,W_{j,n_p})\triangleq\mathcal{L}(\bm{W}_j),~~~\text{in probability}.\label{theorem5.6}
\end{align}
where $W_{j,s}, \widetilde{W}_{j,s}$ with $s=-n_p,...,n_p$ are defined by (\ref{Wtau}) and (\ref{Wtau_para}).
It suffices to show that,
\begin{align}
&\widetilde{\mathcal{L}}\Bigg(\Big(\widetilde{\ell}_{-n_p}(\hat{\bm{\theta}}_{j+1}) -\widetilde{\ell}_{-n_p}(\hat{\bm{\theta}}_{j})\Big),...,\Big(\widetilde{\ell}_{n_p}(\hat{\bm{\theta}}_{j}) -\widetilde{\ell}_{n_p}(\hat{\bm{\theta}}_{j+1})\Big)\Bigg)
~~~\text{based on sample set}~~\{\widetilde{X}_t\}_{t=1}^{2n_p+1}\nonumber\\
&\xrightarrow{d}
\mathcal{L}\Bigg(\Big(\ell_{-n_p}(\bm{\theta}_{j+1}^0) -\ell_{-n_p}(\bm{\theta}_{j}^0)\Big),...,\Big(\ell_{n_p}(\bm{\theta}_{j}^0) -\ell_{n_p}(\bm{\theta}_{j+1}^0)\Big)\Bigg)
~~~\text{based on sample set}~~\{X_t\}_{t=\hat{\tau}_{j-1}^{(3)}+1}^{\hat{\tau}_{j+1}^{(3)}}\nonumber\\
&~~~\text{in probability.}\label{theorem5.3}
\end{align}
Clearly, $g_{t}(\hat{\bm{\theta}}_{j},\hat{\bm{\theta}}_{j+1},\widetilde{\bm{X}}_t)$ defined in Lemma \ref{lemma5} is the extend function of
$\Big(\widetilde{\ell}_{t}(\hat{\bm{\theta}}_{j}) -\widetilde{\ell}_{t}(\hat{\bm{\theta}}_{j+1})\Big)$,
$$g_{t}(\hat{\bm{\theta}}_{j},\hat{\bm{\theta}}_{j+1},\widetilde{\bm{X}}_t) ={\rm sgn}(t)\Big(\widetilde{\ell}_{t}(\hat{\bm{\theta}}_{j}) -\widetilde{\ell}_{t}(\hat{\bm{\theta}}_{j+1})\Big)~\text{for all}~~\widetilde{\bm{X}}_t\in \mathbb{N}^{p+1},$$
and $g_{t}(\bm{\theta}_{j},\bm{\theta}_{j+1},\bm{X}_t)$ fulfills the smoothness condition in Assumption 1 in  \cite{Jentsch2019}. Then combining the Assumption H.\ref{H5} and following the Corollary 3.4 in \cite{Jentsch2019},
there is
\begin{align}
&\widetilde{\mathcal{L}}\Bigg(\Big(\widetilde{\ell}_{-n_p}(\hat{\bm{\theta}}_{j+1}) -\widetilde{\ell}_{-n_p}(\hat{\bm{\theta}}_{j})\Big),...,\Big(\widetilde{\ell}_{n_p}(\hat{\bm{\theta}}_{j}) -\widetilde{\ell}_{n_p}(\hat{\bm{\theta}}_{j+1})\Big)\Bigg)
~~~\text{based on sample set}~~\{\widetilde{X}_t\}_{t=1}^{2n_p+1}\nonumber\\
&\xrightarrow{d}
\mathcal{L}\Bigg(\Big(\ell_{-n_p}(\hat{\bm{\theta}}_{j+1}) -\ell_{-n_p}(\hat{\bm{\theta}}_{j})\Big),...,\Big(\ell_{n_p}(\hat{\bm{\theta}}_{j}) -\ell_{n_p}(\hat{\bm{\theta}}_{j+1})\Big)\Bigg)
~~~\text{based on sample set}~~\{X_t\}_{t=\hat{\tau}_{j-1}^{(3)}+1}^{\hat{\tau}_{j+1}^{(3)}}\nonumber\\
&~~~\text{in probability.}\label{theorem5.4}
\end{align}
Furthermore, since $\ell_{t}(\bm{\theta})$ is continuous with respect to $\bm{\theta}$ at $\bm{\theta}_j^0$ and $\bm{\theta}_{j+1}^0$,
and $\hat{\bm{\theta}}_{j}\xrightarrow{p} \bm{\theta}_j^0$ and $\hat{\bm{\theta}}_{j+1}\xrightarrow{p} \bm{\theta}_{j+1}^0$,
we have
\begin{align}
&\mathcal{L}\Bigg(\Big(\ell_{-n_p}(\hat{\bm{\theta}}_{j+1}) -\ell_{-n_p}(\hat{\bm{\theta}}_{j})\Big),...,\Big(\ell_{n_p}(\hat{\bm{\theta}}_{j}) -\ell_{n_p}(\hat{\bm{\theta}}_{j+1})\Big)\Bigg)
~~~\text{based on sample set}~~\{X_t\}_{t=\hat{\tau}_{j-1}^{(3)}+1}^{\hat{\tau}_{j+1}^{(3)}}\nonumber\\
&\xrightarrow{P}
\mathcal{L}\Bigg(\Big(\ell_{-n_p}(\bm{\theta}_{j+1}^0) -\ell_{-n_p}(\bm{\theta}_{j}^0)\Big),...,\Big(\ell_{n_p}(\bm{\theta}_{j}^0) -\ell_{n_p}(\bm{\theta}_{j+1}^0)\Big)\Bigg)
~~~\text{based on sample set}~~\{X_t\}_{t=\hat{\tau}_{j-1}^{(3)}+1}^{\hat{\tau}_{j+1}^{(3)}}\nonumber\\
&~~~\text{in probability.}\label{theorem5.5}
\end{align}
Combining (\ref{theorem5.4}) and (\ref{theorem5.5}) yields (\ref{theorem5.3}), which implies (\ref{theorem5.6}).
Then, by the $\arg\max$ continuous mapping theorem, for any $n_p\in \mathbb{N}$, there is
\begin{align*}
\widetilde{\tau}_{j,n_p}=\arg\max\limits_{\tau\in\{-n_p,...,n_p\}}\widetilde{W}_{j,\tau}\xrightarrow{d}\arg\max\limits_{\tau}W_{j,\tau}~~\text{in probability},
\end{align*}
and the proof of the part (\ref{theorem5.1}) in Theorem \ref{LRSM5_theory} is completed.

Next, we prove the assertion (\ref{theorem5.2}) in Theorem \ref{LRSM5_theory}.
Note that the sample set $\{{X}_t^{*}\}_{t=1}^{2n_b+1}$ is a concatenation of subsegments $\{X_t\}_{t=\hat{\tau}_{j-1}^{(3)}+1}^{\hat{\tau}_{j}^{(3)}}$ and $\{X_t\}_{t=\hat{\tau}_{j}^{(3)}+1}^{\hat{\tau}_{j+1}^{(3)}}$,
which belong to two stationary process segments in MCP-GCINAR model.
Furthermore, since $\{{X}_t^{*}\}_{t=1}^{n_b+1}$ and $\{{X}_t^{*}\}_{t=n_b+2}^{2n_b+1}$ are independently resampled conditional on the given $\{X_t\}_{t=\hat{\tau}_{j-1}^{(3)}+1}^{\hat{\tau}_{j+1}^{(3)}}$, that is, $\{{X}_t^{*}\}_{t=1}^{n_b+1}$ and $\{{X}_t^{*}\}_{t=n_b+2}^{2n_b+1}$ are independent.
Therefore, there is
\begin{align*}
&\mathcal{L}^{*}(X_{1}^{*},...,X_{2n_b+1}^{*})\xrightarrow{d}\mathcal{L}(X_{\hat{\tau}_{j-1}^{(3)}+1},...,X_{\hat{\tau}_{j+1}^{(3)}})~~~\text{in probability.}
\end{align*}
Furthermore, similar to the proof of the assertion (\ref{theorem5.1}),
we have
\begin{align*}
&\mathcal{L}\Bigg(\Big(\ell_{-n_b}^{*}(\hat{\bm{\theta}}_{j+1}) -\ell_{-n_b}^{*}(\hat{\bm{\theta}}_{j})\Big),...,\Big(\ell_{n_b}^{*}(\hat{\bm{\theta}}_{j}) -\ell_{n_b}^{*}(\hat{\bm{\theta}}_{j+1})\Big)\Bigg)
~~~\text{based on sample set}~~\{{X}_t^{*}\}_{t=1}^{2n_b+1}\nonumber\\
&\xrightarrow{d}
\mathcal{L}\Bigg(\Big(\ell_{-n_b}(\bm{\theta}_{j+1}^0) -\ell_{-n_b}(\bm{\theta}_{j}^0)\Big),...,\Big(\ell_{n_b}(\bm{\theta}_{j}^0) -\ell_{n_p}(\bm{\theta}_{j+1}^0)\Big)\Bigg)
~~~\text{based on sample set}~~\{X_t\}_{t=\hat{\tau}_{j-1}^{(3)}+1}^{\hat{\tau}_{j+1}^{(3)}}\nonumber\\
&~~~\text{in probability.}
\end{align*}
By the $\arg\max$ continuous mapping theorem, for any $n_b\in \mathbb{N}$ and $n_b<\min(\hat{\tau}_{j+1}^{(3)}-\hat{\tau}_{j}^{(3)},\hat{\tau}_{j}^{(3)}-\hat{\tau}_{j-1}^{(3)})$, there is
\begin{align*}
\tau_{j,n_b}^{*}=\arg\max\limits_{\tau\in\{-n_b,...,n_b\}}W_{j,\tau}^{*}\xrightarrow{d}\arg\max\limits_{\tau}W_{j,\tau}~~\text{in probability},
\end{align*}
and the proof of Theorem \ref{LRSM5_theory} is completed.
~\\
\underline{\textbf{Optimal Partitioning Algorithm}}
~\\
~\\
We first define the notations used in the OP Algorithm.\\
Denote the cost function for the $j$th segment by
\begin{align*}
\mathcal{C}(X_{(\tau_{j-1}+1):\tau_j})=-L_{n_j}(\tau_{j-1},\bm{\theta_j},p_j;X_{(\tau_{j-1}+1):\tau_j})+\log(p_j)+\dfrac{p_j+1}{2}\log(n_j).
\end{align*}
where order $p_j$ can be selected through AIC.
\begin{table}[H]
\tabcolsep0.02in
		\begin{tabular}{ll}
			\toprule
\multicolumn{2}{l}{\bf{Optimal Partitioning Algorithm}:}\\
\cmidrule(lr){1-2}
{Input}:& The data set $\{X_t\}_{t=1}^n$, the penalty constant $c=\log(n)$.\\
&The potential change-points set $\hat{\mathcal{J}}^{(1)}$ and $\hat{m}^{(1)}$ from ``First step".\\
{Initialise}:& $F(1)=-c,~\tau^*=[0,\hat{\mathcal{J}}^{(1)},n],~cp(1)=\text{Null}.$\\
Iterate:& for $s_1=1,...,\hat{m}^{(1)}$\\
&~~~~$s_1'=\arg\min_{s_2=1,...,s_1}\big[F(s_2)+\mathcal{C}(X_{(\tau_{s_2}+1):\tau_{s_2+1}})+c\big]$.\\
&~~~~$F(s_1+1)=\min_{s_2=1,...,s_1}\big[F(s_2)+\mathcal{C}(X_{(\tau_{s_2}+1):\tau_{s_2+1}})+c\big]$.\\
&~~~~$\tau'=\tau^*(s_1')$.\\
&~~~~$cp(s_1+1)=\{cp(s_1'),\tau'\}$.\\
&end\\
Output:&the change-points estimates set $\hat{\mathcal{J}}^{(2)}=cp(\hat{m}^{(1)}+2)$.\\
\bottomrule
\end{tabular}
\end{table}
~\\
~\\
\underline{\textbf{Models (B1) - (B9) and (C1) - (C9)}}\\
~\\
Model (B):
\begin{align*}\scriptsize
X_t=\left\{ \begin{array}{lll}
0.5\circ X_{t-1,1}+Z_t&Z_t\overset{i.i.d}{\sim} {\rm Poi} (0.5)&0<t\leq\tau_1^0,\\
0.249\circ X_{t-\tau_1^0-1,2}+0.254\circ X_{t-\tau_1^0-2,2}+0.297\circ X_{t-\tau_1^0-3,2}+Z_t&Z_t\overset{i.i.d}{\sim} {\rm Poi}(1)&\tau_1^0<t\leq\tau_2^0,\\
0.4\circ X_{t-\tau_2^0-1,3}+Z_t&Z_t\overset{i.i.d}{\sim} {\rm Poi}(0.5)&\tau_2^0<t\leq\tau_3^0,\\
0.014\circ X_{t-\tau_3^0-1,4}+0.041\circ X_{t-\tau_3^0-2,4}+0.29\circ X_{t-\tau_3^0-3,4}+0.454\circ X_{t-\tau_3^0-4,4}+Z_t&Z_t\overset{i.i.d}{\sim} {\rm Poi}(2)&\tau_3^0<t\leq\tau_4^0,\\
0.332\circ X_{t-\tau_4^0-1,5}+0.268\circ X_{t-\tau_4^0-2,5}+Z_t&Z_t\overset{i.i.d}{\sim} {\rm Poi}(0.5)&\tau_4^0<t\leq\tau_5^0,\\
0.2\circ X_{t-\tau_5^0-1,6}+Z_t&Z_t\overset{i.i.d}{\sim} {\rm Poi}(4)&\tau_5^0<t\leq\tau_6^0,\\
0.109\circ X_{t-\tau_6^0-1,7}+0.306\circ X_{t-\tau_6^0-2,7}+0.305\circ X_{t-\tau_6^0-3,7}+Z_t&Z_t\overset{i.i.d}{\sim} {\rm Poi}(3)&\tau_6^0<t\leq\tau_7^0,\\
0.3\circ X_{t-\tau_7^0-1,8}+Z_t&Z_t\overset{i.i.d}{\sim} {\rm Poi}(0.5)&\tau_7^0<t\leq\tau_8^0,\\
0.202\circ X_{t-\tau_8^0-1,9}+0.127\circ X_{t-\tau_8^0-2,9}+0.179\circ X_{t-\tau_8^0-3,9}+0.392\circ X_{t-\tau_8^0-4,9}+Z_t&Z_t\overset{i.i.d}{\sim} {\rm Poi}(1)&\tau_8^0<t\leq\tau_9^0,\\
0.3\circ X_{t-\tau_9^0-1,10}+Z_t&Z_t\overset{i.i.d}{\sim} {\rm Poi}(2)&\tau_9^0<t\leq n.
\end{array}\right.
\end{align*}
\textbf{Model (B1)} consists of the first two segments of Model (B) and $\tau_1^0=[0.5n]$. That is
\begin{align*}
X_t=\left\{ \begin{array}{lll}
0.5\circ X_{t-1,1}+Z_t&Z_t\overset{i.i.d}{\sim} {\rm Poi}(0.5)&0<t\leq[0.5n]\\
0.249\circ X_{t-\tau_1^0-1,2}+0.254\circ X_{t-\tau_1^0-2,2}+0.297\circ X_{t-\tau_1^0-3,2}+Z_t&Z_t\overset{i.i.d}{\sim} {\rm Poi}(1)&[0.5n]<t\leq n
\end{array}\right.
\end{align*}
Similarly, we design (B2)-(B9) models as follows.\\
\textbf{Model (B2)} consists of the first three segments of Model (B) and $(\tau_1^0,\tau_2^0)=([0.3n], [0.6n])$.\\
\textbf{Model (B3)} consists of the first four segments of Model (B) and $(\tau_1^0,\tau_2^0,\tau_3^0)=([0.2n], [0.5n], [0.8n])$.\\
\textbf{Model (B4)} consists of the first five segments of Model (B) and

$(\tau_1^0,\tau_2^0,\tau_3^0,\tau_4^0) =([0.2n], [0.4n], [0.6n], [0.8n])$.\\
\textbf{Model (B5)} consists of the first six segments of Model (B) and

$(\tau_1^0,\tau_2^0,\tau_3^0,\tau_4^0,\tau_5^0) = ([0.1n], [0.3n], [0.6n], [0.7n], [0.9n])$.\\
\textbf{Model (B6)} consists of the first seven segments of Model (B) and

$(\tau_1^0,\tau_2^0,\tau_3^0,\tau_4^0,\tau_5^0,\tau_6^0) =([0.1n], [0.2n], [0.3n], [0.5n], [0.8n], [0.9n])$.\\
\textbf{Model (B7)} consists of the first eight segments of Model (B) and

$(\tau_1^0,\tau_2^0,\tau_3^0,\tau_4^0,\tau_5^0,\tau_6^0,\tau_7^0)=([0.1n],[0.2n],[0.3n],[0.4n],[0.5n],[0.8n],[0.9n])$.\\
\textbf{Model (B8)} consists of the first nine segments of Model (B) and

$(\tau_1^0,\tau_2^0,\tau_3^0,\tau_4^0,\tau_5^0,\tau_6^0,\tau_7^0,\tau_8^0)=([0.1n],[0.2n],[0.3n],[0.4n],[0.5n],[0.7n],[0.8n],[0.9n])$.\\
\textbf{Model (B9)} consists of the first ten segments of Model (B) and

$(\tau_1^0,\tau_2^0,\tau_3^0,\tau_4^0,\tau_5^0,\tau_6^0,\tau_7^0,\tau_8^0,\tau_9^0)=([0.1n],[0.2n],[0.3n],[0.4n],[0.5n],[0.6n],[0.7n],[0.8n],[0.9n])$.\\
~\\
Similarly, we set (C1) - (C9) models based on the Model (C).\\
Model (C):
\begin{align*}\scriptsize
X_t=\left\{ \begin{array}{lll}
0.5\ast X_{t-1,1}+Z_t&Z_t\overset{i.i.d}{\sim} {\rm Geo}(1/3)&0<t\leq\tau_1^0\\
0.249\ast X_{t-\tau_1^0-1,2}+0.254\ast  X_{t-\tau_1^0-2,2}+0.297\ast  X_{t-\tau_1^0-3,2}+Z_t&Z_t\overset{i.i.d}{\sim} {\rm Geo}(1/2)&\tau_1^0<t\leq\tau_2^0\\
0.4\ast  X_{t-\tau_2^0-1,3}+Z_t&Z_t\overset{i.i.d}{\sim} {\rm Geo}(1/3)&\tau_2^0<t\leq\tau_3^0\\
0.014\ast  X_{t-\tau_3^0-1,4}+0.041\ast  X_{t-\tau_3^0-2,4}+0.29\ast  X_{t-\tau_3^0-3,4}+0.454\ast  X_{t-\tau_3^0-4,4}+Z_t&Z_t\overset{i.i.d}{\sim} {\rm Geo}(2/3)&\tau_3^0<t\leq\tau_4^0\\
0.332\ast  X_{t-\tau_4^0-1,5}+0.268\ast  X_{t-\tau_4^0-2,5}+Z_t&Z_t\overset{i.i.d}{\sim} {\rm Geo}(1/3)&\tau_4^0<t\leq\tau_5^0\\
0.2\ast  X_{t-\tau_5^0-1,6}+Z_t&Z_t\overset{i.i.d}{\sim} {\rm Geo}(4/5)&\tau_5^0<t\leq\tau_6^0\\
0.109\ast  X_{t-\tau_6^0-1,7}+0.306\ast  X_{t-\tau_6^0-2,7}+0.305\ast  X_{t-\tau_6^0-3,7}+Z_t&Z_t\overset{i.i.d}{\sim} {\rm Geo}(3/4)&\tau_6^0<t\leq\tau_7^0\\
0.3\ast  X_{t-\tau_7^0-1,8}+Z_t&Z_t\overset{i.i.d}{\sim} {\rm Geo}(1/3)&\tau_7^0<t\leq\tau_8^0\\
0.202\ast  X_{t-\tau_8^0-1,9}+0.127\ast  X_{t-\tau_8^0-2,9}+0.179\ast  X_{t-\tau_8^0-3,9}+0.392\ast  X_{t-\tau_8^0-4,9}+Z_t&Z_t\overset{i.i.d}{\sim} {\rm Geo}(1/2)&\tau_8^0<t\leq\tau_9^0\\
0.3\ast  X_{t-\tau_9^0-1,10}+Z_t&Z_t\overset{i.i.d}{\sim} {\rm Geo}(2/3)&\tau_9^0<t\leq n
\end{array}\right.
\end{align*}
\textbf{Model (C1)} consists of the first two segments of Model (B) and $\tau_1^0=0.5n$. That is
\begin{align*}
X_t=\left\{ \begin{array}{lll}
0.5\ast X_{t-1,1}+Z_t&Z_t\overset{i.i.d}{\sim} {\rm Geo}(1/3)&0<t\leq[0.5n]\\
0.249\ast X_{t-\tau_1^0-1,2}+0.254\ast  X_{t-\tau_1^0-2,2}+0.297\ast  X_{t-\tau_1^0-3,2}+Z_t&Z_t\overset{i.i.d}{\sim} {\rm Geo}(1/2)&[0.5n]<t\leq n\\
\end{array}\right.
\end{align*}
Similarly, we design (C2)-(C9) models as follows.\\
\textbf{Model (C2)} consists of the first three segments of Model (B) and $(\tau_1^0,\tau_2^0)=([0.3n], [0.6n])$.\\
\textbf{Model (C3)} consists of the first four segments of Model (B) and $(\tau_1^0,\tau_2^0,\tau_3^0)=([0.2n], [0.5n], [0.8n])$.\\
\textbf{Model (C4)} consists of the first five segments of Model (B) and

$(\tau_1^0,\tau_2^0,\tau_3^0,\tau_4^0) =([0.2n], [0.4n], [0.6n], [0.8n])$.\\
\textbf{Model (C5)} consists of the first six segments of Model (B) and

$(\tau_1^0,\tau_2^0,\tau_3^0,\tau_4^0,\tau_5^0) = ([0.1n], [0.3n], [0.6n], [0.7n], [0.9n])$.\\
\textbf{Model (C6)} consists of the first seven segments of Model (B) and

$(\tau_1^0,\tau_2^0,\tau_3^0,\tau_4^0,\tau_5^0,\tau_6^0) =([0.1n], [0.2n], [0.3n], [0.5n], [0.8n], [0.9n])$.\\
\textbf{Model (C7)} consists of the first eight segments of Model (B) and

$(\tau_1^0,\tau_2^0,\tau_3^0,\tau_4^0,\tau_5^0,\tau_6^0,\tau_7^0)=([0.1n],[0.2n],[0.3n],[0.4n],[0.5n],[0.8n],[0.9n])$.\\
\textbf{Model (C8)} consists of the first nine segments of Model (B) and

$(\tau_1^0,\tau_2^0,\tau_3^0,\tau_4^0,\tau_5^0,\tau_6^0,\tau_7^0,\tau_8^0)=([0.1n],[0.2n],[0.3n],[0.4n],[0.5n],[0.7n],[0.8n],[0.9n])$.\\
\textbf{Model (C9)} consists of the first ten segments of Model (B) and

$(\tau_1^0,\tau_2^0,\tau_3^0,\tau_4^0,\tau_5^0,\tau_6^0,\tau_7^0,\tau_8^0,\tau_9^0)=([0.1n],[0.2n],[0.3n],[0.4n],[0.5n],[0.6n],[0.7n],[0.8n],[0.9n])$.\\

\begin{landscape}
\begin{adjustwidth}{-1cm}{1cm}
The corresponding estimated model based on `0753.HK' data set is as follows:
\begin{align*}\scriptsize
{\rm E}(X_{t}|\mathcal{F}_{t-1})
\left\{
\begin{array}{lr}
0.4220 X_{t-1,1}+0.0404 X_{t-2,1}+0.1558 X_{t-3,1}+5.1320&0<t\leq 420,\\
(0.0737 )~~~~~~~~~~(0.0550 )~~~~~~~~~~(0.0682 )~~~~~~~~~~(1.9947 )\\
0.4392 X_{t-421,2}+0.1215 X_{t-422,2}+0.0945 X_{t-423,2}+13.3739&420<t\leq 994,\\
(0.0763 )~~~~~~~~~~~~(0.0735 )~~~~~~~~~~~~~(0.0556 )~~~~~~~~~~~~~(1.6738 )\\
0.4462 X_{t-995,3}+0.0784 X_{t-996,3}+0.0194 X_{t-997,3}+8.6116&994<t\leq 1934,\\
(0.0496 )~~~~~~~~~~~~(0.0386 )~~~~~~~~~~~~~(0.0307 )~~~~~~~~~~~~~(1.0422 )\\
0.4475 X_{t-1935,4}+0.0872 X_{t-1936,4}+0.0667 X_{t-1937,4}+5.1398&1934<t\leq 3175,\\
(0.0404 )~~~~~~~~~~~~~~(0.0406 )~~~~~~~~~~~~~~(0.0309 )~~~~~~~~~~~~~~(0.6315 )\\
0.3395 X_{t-3176,5}+0.0882 X_{t-3177,5}+0.1140 X_{t-3178,5}+8.6316&3175<t\leq 3701,\\
(0.0475 )~~~~~~~~~~~~~~(0.0456 )~~~~~~~~~~~~~~(0.0477 )~~~~~~~~~~~~~~(0.9677 )\\
0.3543 X_{t-3702,6}+0.0408 X_{t-3703,6}+13.6244&3701<t\leq 4030,\\
(0.0565 )~~~~~~~~~~~~~~(0.0609 )~~~~~~~~~~~~~~(1.5216 )\\
0.3781 X_{t-4031,7}+0.0816 X_{t-4032,7}+0.1640 X_{t-4033,7}+5.7080&4030<t\leq 4550.\\
(0.0502 )~~~~~~~~~~~~~~(0.0534 )~~~~~~~~~~~~~~(0.0528 )~~~~~~~~~~~~~~(0.9627 )\\
\end{array}\right.
\end{align*}
\begin{figure}[H]
\includegraphics[width=10.5in,height=3in]{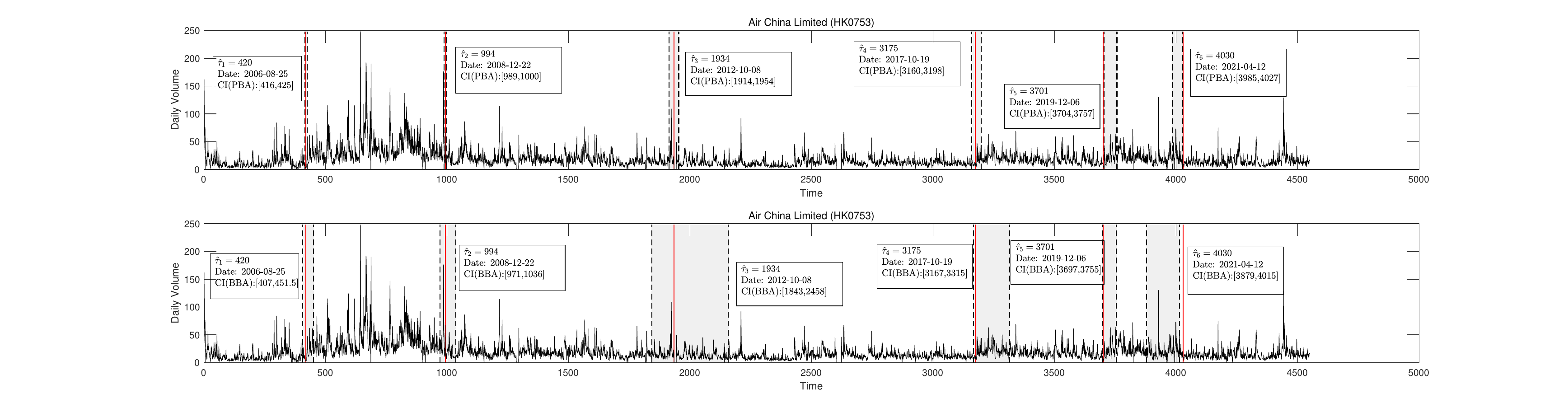}
\vspace{-6mm}
\caption{Sample path of the 0753.HK daily trades volume and estimate results given by the LRSM.\protect\\ The red lines are the LRSM change-points estimates and the shaded regions are the corresponding PBA and BBA CIs.}\label{box_r_known}
\end{figure}
\end{adjustwidth}
\end{landscape}
\begin{landscape}
\begin{adjustwidth}{0.3cm}{3cm}
The corresponding estimated model based on `LHA.DE' data set is as follows:
\begin{align*}\scriptsize
{\rm E}(X_{t}|\mathcal{F}_{t-1})=
\left\{
\begin{array}{lr}
0.2718 X_{t-1,1}+0.1389 X_{t-2,1}+0.0905 X_{t-3,1}+0.0554 X_{t-4,1}+0.0761 X_{t-5,1}+0.6157&0<t\leq 1159,\\
(0.0436 )~~~~~~~~~~(0.0365 )~~~~~~~~~~(0.0286 )~~~~~~~~~~(0.0338 )~~~~~~~~~~~(0.0318 )~~~~~~~~~~(0.0650 )\\
0.3542 X_{t-1160,2}+0.1026 X_{t-1161,2}+0.0663 X_{t-1162,2}+0.0579 X_{t-1163,2}+0.0848 X_{t-1164,2}+0.0226 X_{t-1165,2}+0.0434 X_{t-1166,2}+1.2173&1159<t\leq 5886,\\
(0.0174 )~~~~~~~~~~~~~(0.0167 )~~~~~~~~~~~~~~~(0.0161 )~~~~~~~~~~~~~~(0.0151 )~~~~~~~~~~~~~(0.0154 )~~~~~~~~~~~~~~~~(0.0152 )~~~~~~~~~~~~~~(0.0138 )~~~~~~~~~~~~~(0.0848 )\\
0.4655 X_{t-5887,3}+0.2113 X_{t-5888,3}+0.0095 X_{t-5889,3}+0.0172 X_{t-5890,3}+0.0356 X_{t-5891,3}+0.1213 X_{t-5892,3}+1.1921&5886<t\leq 6495,\\
(0.0597 )~~~~~~~~~~~~~(0.0610 )~~~~~~~~~~~~~~~(0.0421 )~~~~~~~~~~~~~~(0.0384 )~~~~~~~~~~~~~~(0.0647 )~~~~~~~~~~~~~~~(0.0467 )~~~~~~~~~~~~~~(0.2632 )\\
0.1988 X_{t-6496,4}+0.1343 X_{t-6497,4}+3.7911&6495<t\leq 6774.\\
(0.0686 )~~~~~~~~~~~~~(0.0552 )~~~~~~~~~~~~~~~(0.4654)\\
\end{array}\right.
\end{align*}
\begin{figure}[H]
\includegraphics[width=10.5in,height=3in]{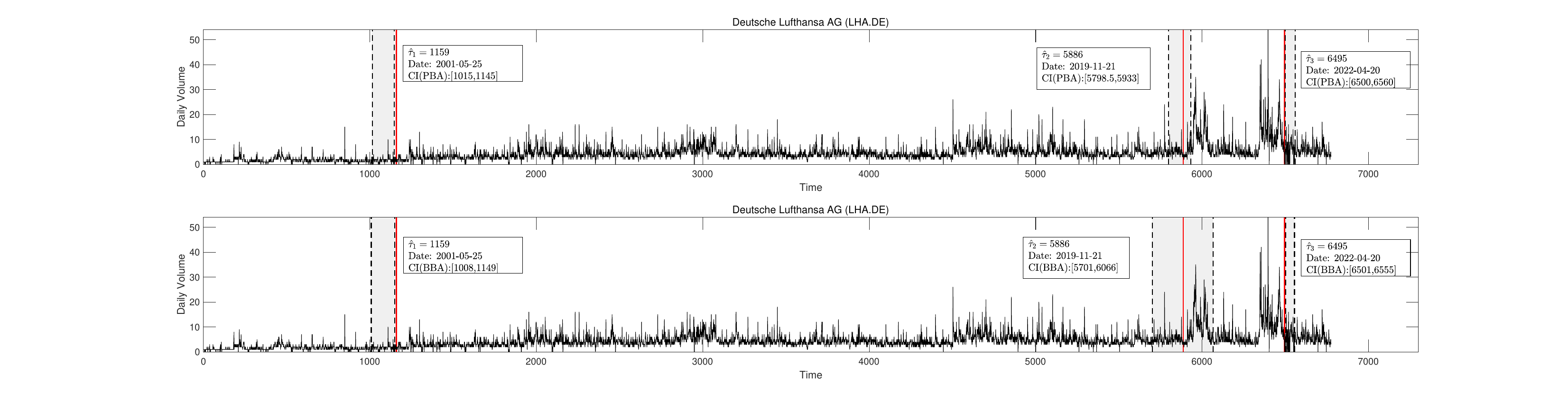}
\vspace{-6mm}
\caption{Sample path of the LHA.DE daily trades volume and estimate results given by the LRSM.\protect\\ The red lines are the LRSM change-points estimates and the shaded regions are the corresponding PBA and BBA CIs.}
\end{figure}
\end{adjustwidth}
\end{landscape}
\begin{landscape}
\begin{adjustwidth}{-1cm}{1cm}
The corresponding estimated model based on `UAL' data set is as follows:
\begin{align*}\scriptsize
{\rm E}(X_{t}|\mathcal{F}_{t-1})
\left\{
\begin{array}{lr}
0.4411 X_{t-1,1}+0.0410 X_{t-2,1}+0.0895 X_{t-3,1}+0.0803 X_{t-4,1}+0.1100 X_{t-5,1}+0.0179 X_{t-6,1}+0.0830 X_{t-7,1}+0.8574&0<t\leq 3480,\\
(0.0344 )~~~~~~~~~~(0.0219 )~~~~~~~~~~(0.0252 )~~~~~~~~~~(0.0210 )~~~~~~~~~~~(0.0235 )~~~~~~~~~~(0.0213 )~~~~~~~~~~~(0.0202 )~~~~~~~~~~(0.1057 )\\
0.4552 X_{t-3481,2}+0.0339 X_{t-3482,2}+0.1179 X_{t-3483,2}+0.1596 X_{t-3484,2}+0.0392 X_{t-3485,2}+0.1522 X_{t-3486,2}+0.7725&3480<t\leq 4338.\\
(0.0521 )~~~~~~~~~~~~~(0.0405 )~~~~~~~~~~~~~~~(0.0430 )~~~~~~~~~~~~~~(0.0486 )~~~~~~~~~~~~~~(0.0435 )~~~~~~~~~~~~~~(0.0461 )~~~~~~~~~~~~~~(0.2826 )\\
\end{array}\right.
\end{align*}
\begin{figure}[H]
\includegraphics[width=10.5in,height=3in]{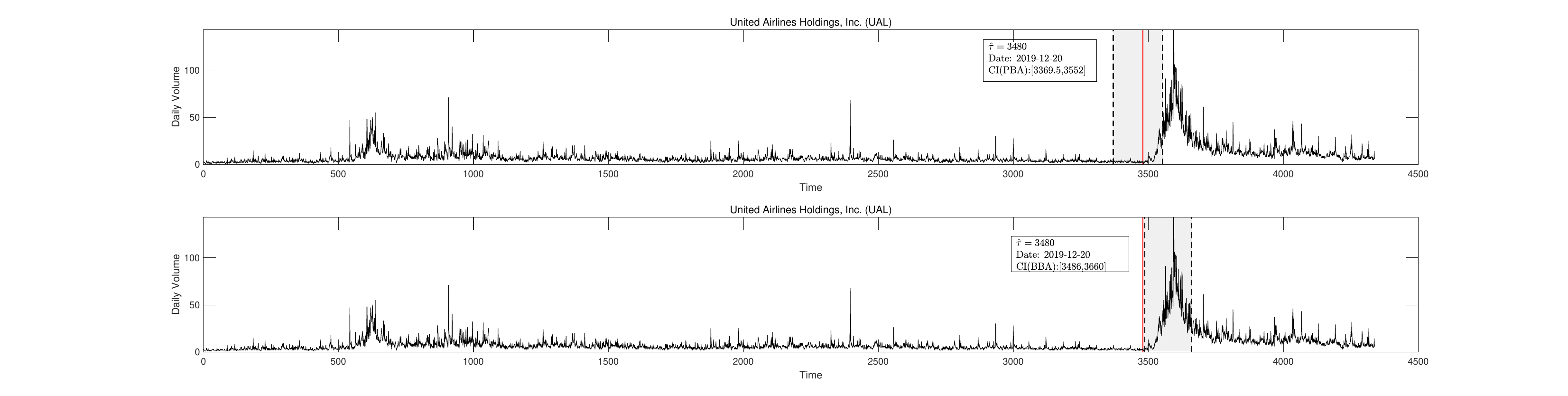}
\vspace{-6mm}
\caption{Sample path of the UAL daily trades volume and estimate results given by the LRSM.\protect\\ The red line is the LRSM change-points estimate and the shaded regions are the corresponding PBA and BBA CIs.}
\end{figure}
\end{adjustwidth}
\end{landscape}

\end{document}